\documentclass[twocolumn,pra,superscriptaddress]{revtex4-2}

\usepackage{graphicx}  
\usepackage{amsmath}
\usepackage{dcolumn}
\usepackage{bm}
\usepackage{footmisc}

\begin{document}

\preprint{AIP/123-QED}

\title{
Expansion of strongly interacting dipolar bosons in 1D optical lattices}
\author{Rhombik Roy }
\thanks{Currently at: Department of Physics, University of Haifa, 3498838 Haifa, Israel.}
\email{rroy@campus.haifa.ac.il}
\affiliation{Department of Physics, Presidency University, 86/1   College Street, Kolkata 700073, India.}

\author{Andrea Trombettoni }
\email{andreatr@sissa.it}
\affiliation{Department of Physics, University of Trieste, Strada Costiera 11, I-34151 Trieste, Italy.}
\affiliation{CNR-IOM DEMOCRITOS Simulation Centre and SISSA, Via Bonomea 265, I-34136 Trieste, Italy.}

\author{Barnali Chakrabarti}
\email{bchakrab@ictp.it}
\affiliation{Department of Physics, Presidency University, 86/1   College Street, Kolkata 700073, India.}
\affiliation{The Abdus Salam International Center for Theoretical Physics, 34100 Trieste, Italy.}

\date{\today}

\begin{abstract}
We numerically study the expansion dynamics of initially localized dipolar bosons in a homogeneous 1D optical lattice for different initial states. Comparison is made to interacting bosons with contact interaction. For shallow lattices the expansion is unimodal and ballistic, while strong lattices suppress tunneling. However for intermediate lattice depths a strong interplay between dipolar interaction and lattice depth occurs. The expansion is found to be bimodal, the central cloud expansion can be distinguished from the outer halo structure. In the regime of strongly interactions dipolar bosons exhibit two  time scales, with an initial diffusion and then arrested transport in the long time; while strongly interacting bosons in the fermionized limit exhibit ballistic expansion. Our study highlights how different lattice depths and initial states can be manipulated to control tunneling dynamics.
\end{abstract}

\maketitle

\section{Introduction} \label{intro}
Recent experimental developments allow a very precise control of properties of ultracold atomic gases in optical lattices. 
This tunability 
makes the nonequilibrium dynamics of strongly correlated many-body systems as one of the most challenging problems
for theoretical physics~\cite{RevModPhys.80.885,RevModPhys.83.863}. External optical lattices can be superimposed 
with typical experimental time-scales 
for the dymaics faster than typical relaxation and decoherence rates. 
Thus, ultracold atoms offer the possibility to study the transport properties and out-of-equilibrium phenomena in well controlled way --- number of atoms, inter-atomic interaction, lattice depth all being 
controllable parameters. In the understanding of non-equilibrium dynamics, special interest has been put for expansion dynamics of strongly correlated bosons in 1D optical lattice~\cite{Schneider,Ronzheimer,Collura,science.305,Wang,Mistakidis,PhysRevA.97.053626}. The two prototypical transport mechanisms, ballistic and diffusive transport, are now well confirmed in experiments.

The effect of interaction on expansion dynamics of confined bosonic and fermionic gases on 1D and 2D latticed has been extensively studied experimentally~\cite{Schneider,Ronzheimer,Langer,Weiss,Weiss1} as well theoretically~\cite{Weixu,Anna,Loris,Wolf,Juan,Langer1,Frank}. 
Most of the studies in this direction considered the Mott insulator state as the initial state and various aspects of the time dynamics 
after an abrupt reduction of lattice height have been considered. 

A very interesting experimental observation is the bimodal structure in the expansion dynamics--diffusive dynamics in the center is surrounded by the ballistic wings~\cite{Ronzheimer}. 
The interplay between the interaction and dimensionality has been also the subject of considerable interest. 
Self trapping in the array of $^{87}Rb$ atoms in 1D and quasi-1D tubes are also studied when the mean-field energy gradient suppresses the tunneling and thus 
stops the expansion.  The self-trapping phenomenon, 
an effect induced by the nonlinear interaction of the condensate, has been first experimentally observed in ~\cite{Anker} for the expansion in an optical lattice and in~\cite{Albiez05} for a double well potential. The self-trapped phenomenon has been extensively studied in nondipolar BECs~\cite{Smerzi97,Raghavan99,Trombettoni01,Trombettoni_2001,Wang,Creffield,Bao} and for dipolar bosons in the double well potential~\cite{Bo,rhombik_epjd}, see more references in \cite{Oberthaler06}. 

In a typical experimental setup, atoms are first loaded in a 1D optical lattice with an additional dipole trap, which forms an array of 2D pancake.like BECs. Then the dipole trap is suddenly removed allowing the free expansion of the bosons in the homogeneous lattice. 
In most of the studies of the temporal expansion of clouds in optical lattices, the on-site interaction is the most common assumption. However,  
it has been shown that the nearest neighbour coupling rate is not negligible compared to atom tunneling rate near a broad Feshbach resonance~\cite{Duan}. Rydberg dressing also gives rise to on-site as well as long-range interactions~\cite{Lama}, see the recent review~\cite{Defenu23}. Using Ramsey spectroscopy quantum dynamics of disordered dipolar interacting ultracold molecules in a partially filled optical lattice has also been studied~\cite{Kaden}. 
Using the Feshbach resonance technique,  it is now possible to 
considerably reduce the effects of short-range interactions and to
study the 
dominating effect introduced by
the dipolar interaction~\cite{Koch,Koch1,Koch2}. This facilitates additional control
parameter to manipulate the tunneling dynamics in
optical lattices. Thus rapid experimental advances in controlling ultracold
atomic interactions make it possible to
realize enticing states specially in reduced dimension. 

Dipolar ultracold atoms have also attracted much interest which is corroborated by experimental realization of dipolar Bose Einstein condensate consists of chromium~\cite{cro}, dysprosium~\cite{dys} and erbium atoms~\cite{erb}. The  
non-local dipole-dipole interaction and its anisotropic nature leads to rich and exotic many-body physics which are distinctly different from the BEC with contact interaction~\cite{ref1, ref2}. In the recent experiment~\cite{Lin:2023}, the dipolar interaction is tuned which makes it possible to realize novel quantum phases in the strongly correlated lattice with long-range interaction. 
Dipolar interactions in reduced dimensions facilitate
the exploration of fascinating many-body phenomena. One such phenomenon is the crystallization in one- and two-dimensional systems~\cite{ref3,ref4,ref5,sangita_sci.rep}. Crystallization is the consequence of strong repulsive long-ranged tail of the dipolar interaction when the bosons exhibit maximal separation. The corresponding counterpart with contact interaction is the fermionization when the strongly interacting bosons also escape their spatial overlap.  
Crystallized bosons can be distinguished from the fermionized bosons from the peak-to-peak separation in the density distribution~\cite{sangita_sci.rep}.

In the present manuscript we aim at describing the expansion dynamics as 
to the one in~\cite{Ronzheimer}, but here with dipolar interaction, to study the expansion dynamics of initially localized 
interacting dipolar bosons in 1D optical lattices. 
We will compare our results with the ones for the expansion 
in 1D optical lattices of bosons interacting via short-range interactions.
We consider 
diverse initial setups to unravel the intriguing effect of interaction in the expansion and correlation dynamics.  
Our system is prepared in the ground state of the many-body Hamiltonian. 
We then impose a harmonic trap such that the bosons 
localize in the 
central wells. The choice of interaction strength both for contact and dipolar interaction guarantees that without harmonic trap localization the initial state is fermionized in the first case and crystallized for the second case. The interaction strength is kept fixed throughout the analysis. We have made scanning in the weak and intermediate interaction strength, however the 
more interesting many-body features in the dynamics are observed in the extreme limit of fermionized and crystallized phases. We
present numerically exact many-body expansion dynamics by solving the time-dependent many-boson Schr\"odinger equation utilizing the multiconfigurational time-dependent
Hartree method for bosons (MCTDHB)~\cite{Streltsov:2006,Streltsov:2007,Alon:2007,Alon:2008,Lode:2016,Fasshauer:2016,Lode:2020}, implemented in the MCTDH-X software~\cite{Lin:2020,MCTDHX}. We extract key measures such as one-body density dynamics and Glauber correlation functions. We analyse results for $N=4$ interacting bosons in $S=21$ lattices. We make a 
scan in lattice depth in the range of 
$\{ V_0=0.1E_r, 10.0E_r \}$, $V_0$ being the lattice strength and $E_r$ the lattice recoil energy, and discuss 
four representative cases having different initial set ups which lead to four 
distinct expansion mechanisms. At time $t=0$, we instantaneously switch off the harmonic trap and allow the bosons to expand. Sudden removal of the harmonic trap should ballistically favor the tunneling into the outer wells, however the tunneling dynamics is strongly influenced by the 
non-local interaction of dipolar bosons. For very weak lattices, $(V_0= 0.1 E_r)$, when one has 
effectively a harmonic trap, we observe ballistic expansion in both cases but with distinct many-body features. When the fermionized bosons loose their independent jet-like structure quickly, the crystallized bosons carry the many-body features for long time until they reach the lattice boundary. On the other hand, very strong lattices, $(V_0=10.0 E_r)$ inhibits the tunneling. When the fermionized bosons remain localized for quite long time and then start tunneling in the adjacent wells; correspondigly, the fermionized 
bosons remain localized for entire region of dynamics. Interplay between the lattice potential and strong interparticle interaction 
controls the tunneling dynamics for intermediate lattice depth potentials: $V_0=1.0 E_r$ and $V_0= 5.0 E_r$. We are able to observe dominating effect of 
non-local interaction. When the bosons with strong contact interaction exhibits ballistic expansion, dipolar bosons exhibit diffusive expansion in short time and arrested transport at longer time. For intermediate lattice depths, the expansion is bimodal; the expansion of high-density central cloud can be distinguished from the incoherent outer cloud expansion with very intriguing many-body features. 

Self trapping is a well understood phenomenon in the context of non-dipolar bosons and is caused due to competition between the lattice depth and mean-field energy gradient. When the nonlinear effect suppresses the tunneling the expansion is inhibited and causes localization of wave packets. As the root cause of self trapping is the interatomic interaction, it is inherently a many body phenomena, and mean-field Gross-Pitaevskii equation is commonly used to study the self-trapping in three dimensions. However, in 1D gases, strong quantum correlations are built up and strongly interacting bosons do not 
macroscopically occupy single-particle wave function. The many-body wave function is now a set of spatially distinct single particle wave function in each lattice to avoid interaction. For dipolar bosons, there will be an 
nterplay between lattice depth, strength of dipolar interaction as well as long range correlation arising from long range part of dipolar interaction. The scenario is now more complicated --- the tunneling should be favored due to long-range repulsion of the bosons as well as due to sudden removal of the harmonic trap. However, due to very strong quantum correlation for dipolar bosons, tunneling is 
quickly suppressed. There are 
strong mismatched correlations between adjacent sites as the atoms are already well isolated due to non-local interaction. To allow tunneling an atom needs to pay large interaction energy to adjust this mismatched correlations. Thus dipolar bosons favor localization instead of quantum tunneling which is described as {\it{arrested transport}} in the manuscript.

The paper is organized as follows. In Sec. II, we introduce the theoretical model, while in Sec III introduces the quantities of interest. In Sec IV, we present the initial state and the setup. Sec V presents the quench dynamics for four different initial set ups distributed over four subsections. Sec VI is devoted to our conclusions and our final remarks.

\section{The model
}\label{numerics}
The equation of motion for $N$ interacting bosons in a time-dependent system is governed by the time-dependent Schr\"odinger equation as
$\hat{H} \psi = i \frac{\partial \psi}{\partial t}$ (with $\hbar=1$). The total Hamiltonian for the system takes the following form
\begin{equation} 
\hat{H}(x_1,x_2, \dots x_N)= \sum_{i=1}^{N} \hat{h}(x_i) + \sum_{i<j=1}^{N}\hat{W}(x_i - x_j).
\label{propagation_eq}
\end{equation}
The Hamiltonian $\hat{H}$ will be expressed in dimensionless units, see below in Section IV. $\hat{h}(x) = \hat{T}(x) + \hat{V}_{trap}(x)$ is the one-body Hamiltonian. $\hat{T}(x)$ is the kinetic energy operator and $\hat{V}_{trap}(x)$ is the external trapping potential. $\hat{W}(x_i - x_j)$ is the two-body interaction. $x_i$ is the coordinate of the $i$- th boson. The form of $\hat{V}_{trap}(x)$ and $\hat{W}(x_i - x_j)$ will be specified in Section IV. 

To solve the  time-dependent  many-boson
Schr\"odinger, we have deployed an in-principle numerically exact, many-body,  method, the 
MCTDHB, whose  
equation of motions can be found in~\cite{Streltsov:2006,Streltsov:2007,Alon:2007,Alon:2008}. We provide a brief description of the implementation of MCTDHB in Appendix A.



\section{Quantities of interest}\label{quant}
To explore the 
expansion dynamics in optical lattices of dipolar bosons, we employ various measures to characterize the system's behavior. These measures provide insights into the spatial coherence and correlations within the evolving bosonic cloud.

(i) The reduced one-body density matrix in coordinate space, denoted as $\rho^{(1)}(x_{1}^{\prime}\vert x_{1};t)$, is a fundamental quantity of interest. It captures the density distribution of the system by integrating over the remaining coordinates, defined as:
\begin{equation}
\begin{split}
\rho^{(1)}(x_{1}^{\prime}\vert x_{1};t) = N\int dx_{2},dx_{3}...dx_{N} \\ \psi^{*}(x_{1}^{\prime},x_{2},\dots,x_{N};t) \psi(x_{1},x_{2},\dots,x_{N};t),
\label{onebodydensity}
\end{split}
\end{equation}
where $N$ represents the total number of particles. It is diagonal part gives the one-body density $\rho(x,t)$.

(ii) The $p$-th order reduced density matrix in coordinate space, denoted as $\rho^{(p)}(x_{1}^{\prime}, \dots, x_{p}^{\prime} \vert x_{1}, \dots, x_{p};t)$. This quantity describes the joint probability distribution of finding $p$ particles at specific positions and defined as
\begin{equation}
\begin{split}
\rho^{(p)}(x_{1}^{\prime}, \dots, x_{p}^{\prime} \vert x_{1}, \dots, x_{p};t) = \frac{N!}{(N-p)!}\int dx_{p+1},...dx_{N} \\ \psi^{*}(x_{1}^{\prime},\dots, x_{p}^{\prime},x_{p+1},\dots,x_{N};t) \psi(x_{1},\dots,x_{p},x_{p+1}\dots,x_{N};t).
\label{pbodydensity}
\end{split}
\end{equation}

(iii) The $p$-th order Glauber correlation function $g^{(p)}(x_{1}^{\prime}, \dots, x_{p}^{\prime}, x_{1}, \dots, x_{p};t)$ measures degree of spatial coherence and
correlations within the system. It is defined as:
\begin{equation}
\begin{split}
g^{(p)}(x_{1}^{\prime}, \dots, x_{p}^{\prime}, x_{1}, \dots, x_{p};t) = \\ \frac{\rho^{(p)}(x_{1}, \dots, x_{p} \vert x_{1}^{\prime}, \dots, x_{p}^{\prime};t)}{\sqrt{\prod_{i=1}^{p} \rho^{(1)} (x_i \vert x_i;t) \rho^{(1)} (x_{i}^{\prime} \vert x_{i}^{\prime};t)}}.
\label{correlation}
\end{split}
\end{equation}
Similarly, the diagonal elements of $g^{(p)}(x_{1}^{\prime}, \dots, x_{p}^{\prime}, x_{1}, \dots, x_{p};t)$, denoted as $g^{(p)}( x_{1}, \dots, x_{p};t)$, provide a measure of $p$-th order coherence. 
If $\vert g^{(p)}( x_{1}, \dots, x_{p};t) \vert = 1$, the system is fully coherent, while deviations from unity indicate partial coherence. Specifically, $g^{(p)}( x_{1}, \dots, x_{p};t) > 1$ implies correlated detection probabilities at positions $x_{1}, \dots, x_{p}$, while $g^{(p)}( x_{1}, \dots, x_{p};t) < 1$ indicates anti-correlations.
These measures collectively provide valuable insights into the evolving density profile, correlations, and coherence of the bosonic system as it undergoes expansion in the shallow and deep optical lattices. 
\begin{figure}
\centering
\includegraphics[scale=0.19, angle=-0]{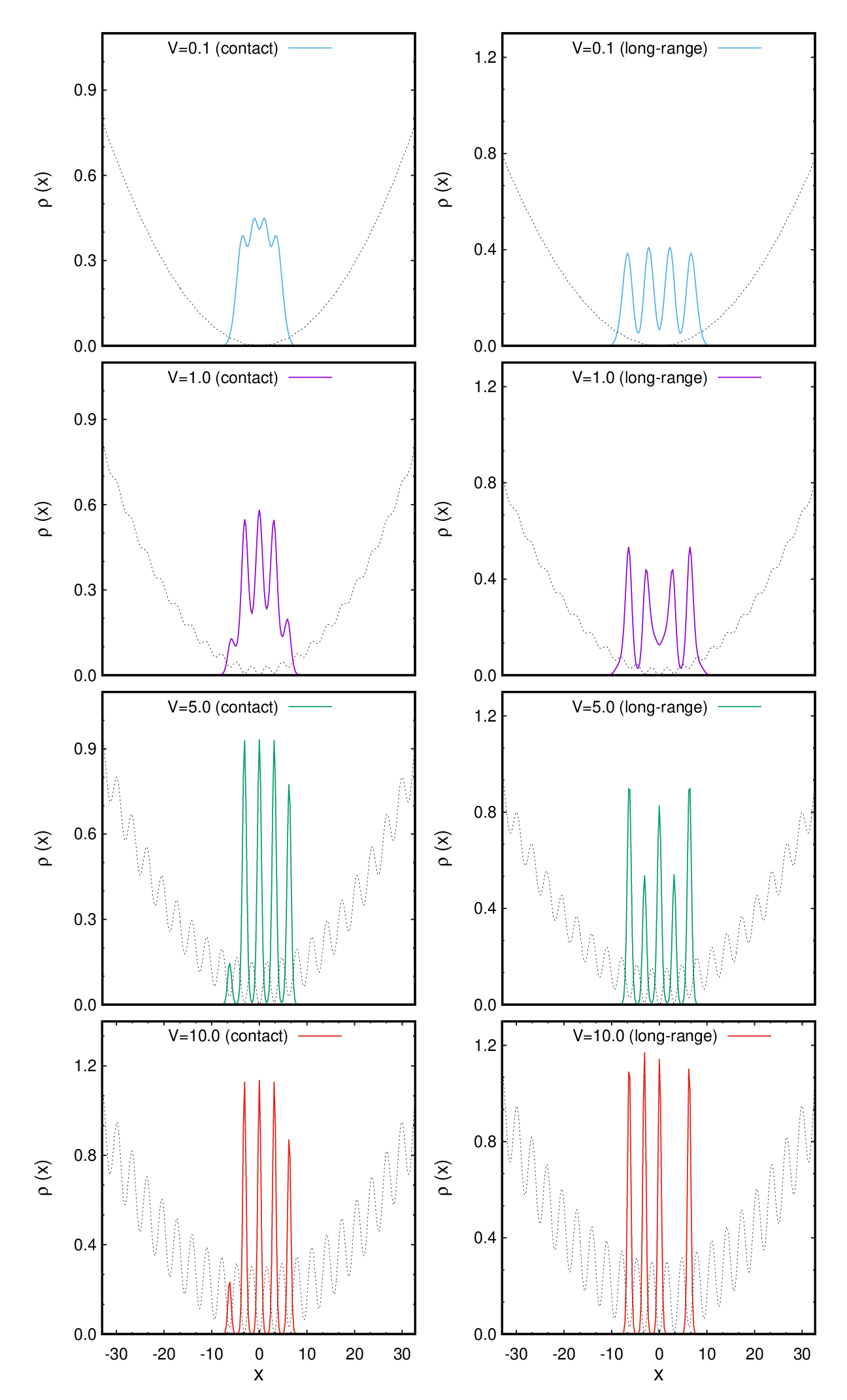}
\caption{Initial density profiles for contact and dipolar interactions in the 
Hamiltonian of Eq. (5) for four bosons in $21$  lattice sites of various lattice depths, $V = 0.1E_r, 1.0E_r, 5.0E_r, 10.0E_r$ from top to bottom. Left panel is for contact interaction ($\lambda = 25$). Right panel is for dipolar interaction ($g_d = 25$). See the text for details. All quantities are dimensionless.}
\label{fig1}
\end{figure}


\section{Initial State}\label{initial}
We prepare the initial state in a combined harmonic oscillator trap of frequency $\omega$ and a 1D 
optical lattice. The many-body Hamiltonian reads as 
\begin{equation}
    \hat{H}= \sum_{i=1}^N \left( - \frac{1}{2}\frac{\partial^2}{\partial x_i^2} + \frac{1}{2} \omega^2 x_i^2 + V sin^2 (k x_i)\right) + \sum_{i<j}^N \hat{W} (x_i - x_j)
\end{equation}
$V$ is the depth of the optical lattice, $k$ is the wave vector and $l =\frac{\pi}{k}$ is the periodicity of the optical lattice. The many-body Hamiltonian is rescaled in the unit of recoil energy $E_r = \frac{\hbar^2 k^2}{2 m}$. Thus the unit of time, length, and interaction strength are $\hbar E_r^{-1}$, $k^{-1}$ and $2 E_r k^{-1}$ respectively. The spatial extension of the optical lattice is $x_+ = 10\pi$ to $x_- = -10\pi$ to restrict the number of lattice sites to $21$. We impose the hard wall boundary condition. The Hamiltonian can be written in dimensionless form by dividing the dimensionful Hamiltonian by $\frac{\hbar^{2}}{mL^2}$, $L$ is the arbitrary length scale which leads to $\hbar$= $m$= $k$ =1.  The initial state is the ground state of the many-body Hamiltonian with harmonic trap frequency $\omega = 0.22$ and lattice depth is chosen as $V = 0.1 E_r$, $V = 1.0 E_r$, $V =5.0 E_r$ and $V = 10.0 E_r$ to study the dynamics in the entire range of lattice height which can be controlled experimentally.\\
The harmonic trap leads to localization towards the central wells of the lattice. To ensure that the system is in quasi-1D regime, we assume strong transverse confinement. The two-body interaction for contact interaction reads as
\begin{equation}
    \hat{W} (x_i - x_j) = \lambda \delta (x_i - x_j)
\end{equation}

where $\lambda$ is the interaction strength determined by the scattering length $a_s$  and the transverse confinement frequency. For 
dipolar interaction,
\begin{equation}
    \hat{W} (x_i - x_j) =\frac{g_d}{\vert (x_i - x_j) \vert ^3 + \alpha}
\end{equation}
$g_d$ is the pure dipolar interaction strength; $g_d = \frac{d_m^{2}}{4\pi\epsilon_0}$for electric
dipoles and $g_d = \frac{d_m^2\mu_0} {4\pi}$ for magnetic dipoles, $d_m$ being
the dipole moment, $\epsilon_0$ is the vacuum permittivity, and $\mu_0$ is the
vacuum permeability. $\alpha$ is the short-range cutoff to avoid singularity at $x_i = x_j$. We choose $\alpha$ =  0.05, which corresponds to $a_{\perp} = 0.37$  and an aspect ratio = 42.5~\cite{Santos,Deuretzbacher}. In general, the dipole-dipole interaction potential in 1D
also includes a contact term owing to the
transverse confinement, however that can be safely neglected for
strong interaction strengths~\cite{Budhaditya}. 

We consider expansion dynamics of $N=4$ strongly interacting bosons interacting via contact interaction of strength ($\lambda =25.0$) and dipolar interaction strength ($g_d =25.0$). However, the physics remains same for $N=3$ and $N=5$. We restrict to the smaller number of particles to ensure convergence in the observed many-body dynamics. 
We solve the set of coupled equations using MCTDH-X software (R-MCTDHB package) ~\cite{Lin:2020,MCTDHX}. For the relaxed state, propagation is done in imaginary time with an initial guess which converges to the ground state of the system. 

In Fig.~\ref{fig1} , we plot the initial state  both for contact and dipolar interaction for four different choices of lattice depth. 
For $V=0.1E_r$, when the effective trap is almost a HO trap, the one-body density for $\lambda=25$, is the fermionic density distribution. 
The emergence of four maxima corresponds to four strongly interacting bosons in the fermionized limit. The two innermost humps are pronounced at the center of the trap where the effective potential is zero, the two outermost humps are less pronounced due to the larger distance from the center of the trap. The corresponding one-body density for $g_d=25$, exhibits a crystal phase. The four humps again signify the four strongly interacting dipolar bosons, this is the similarity between a crystal phase and a fermionized phase, in both cases the strongly interacting bosons escape their spatial overlap. However, unlike the contact interaction where the humps were "non isolated", in the crystal phase they are "isolated" due to the long range tail of the dipolar interaction.  For crystallized dipolar bosons, the value of the density at the minima between the humps tends to zero and the spreading of the density profile is broadened. Thus the density modulation in the crystallized phase is significantly different from the fermionized phase. Now increasing the lattice depth, the HO trap and lattice height interplay and the effective lattice structure becomes inhomogeneous. For $V=1.0E_r$, with contact interaction, three distinct humps are distributed over three lattice sites around the central lattice and two weak humps are observed in the outer lattice. Whereas  the initial state is less significantly affected by the dipolar interaction for this weak lattice. Four distinct humps with some modulation in the density are observed as before. For much higher lattice depth, $V=5.0E_r$, four tightly trapped and completely isolated peaks around the central lattice with a very weak outer peak are observed in the density profile for contact interaction. For dipolar interaction, we observe five tightly bound peaks distributed over five consecutive lattices around the central lattice. It is to be noted that the weak outer peak in the density profile for contact interaction and  the strong outer peaks for dipolar interaction will play significant role in the tunneling dynamics. For lattice depth $V=10.0E_r$,  there is no further effect on density modulation for contact interaction; only the peaks become more strongly bound due to larger height of lattice. Whereas density pattern for dipolar bosons are strongly affected. We observe four distinct peaks in the density distribution but long range interaction strongly interplay with lattice depth and one peak settles to a far right lattice site escaping a near one.


\section{Expansion dynamics in the lattice}
\begin{figure}
\centering
\includegraphics[scale=0.2, angle=-90]{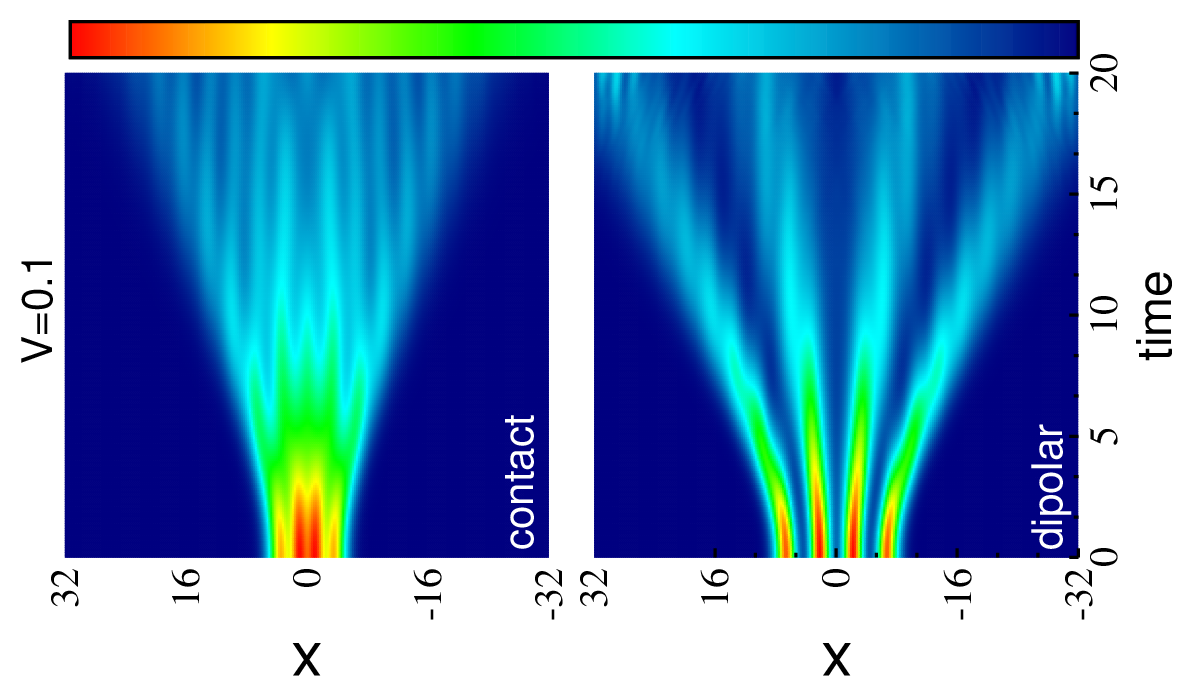}
\caption{ Density dynamics on sudden removal of trap in the Hamiltonian (Eq.(5)), for $N=4$ bosons, orbital $M=15$, lattice sites $S=21$, lattice depth $V=0.1 E_r$. The upper panel depicts post quench density dynamics for contact interaction ($\lambda=25$), while the lower panel shows the same for dipolar interaction ($g_d=25$). 
}
\label{fig2}
\end{figure}
\subsection{Expansion dynamics in weak lattice: $V=0.1 E_r$}
\begin{figure}
\centering
\includegraphics[scale=0.3, angle=0]{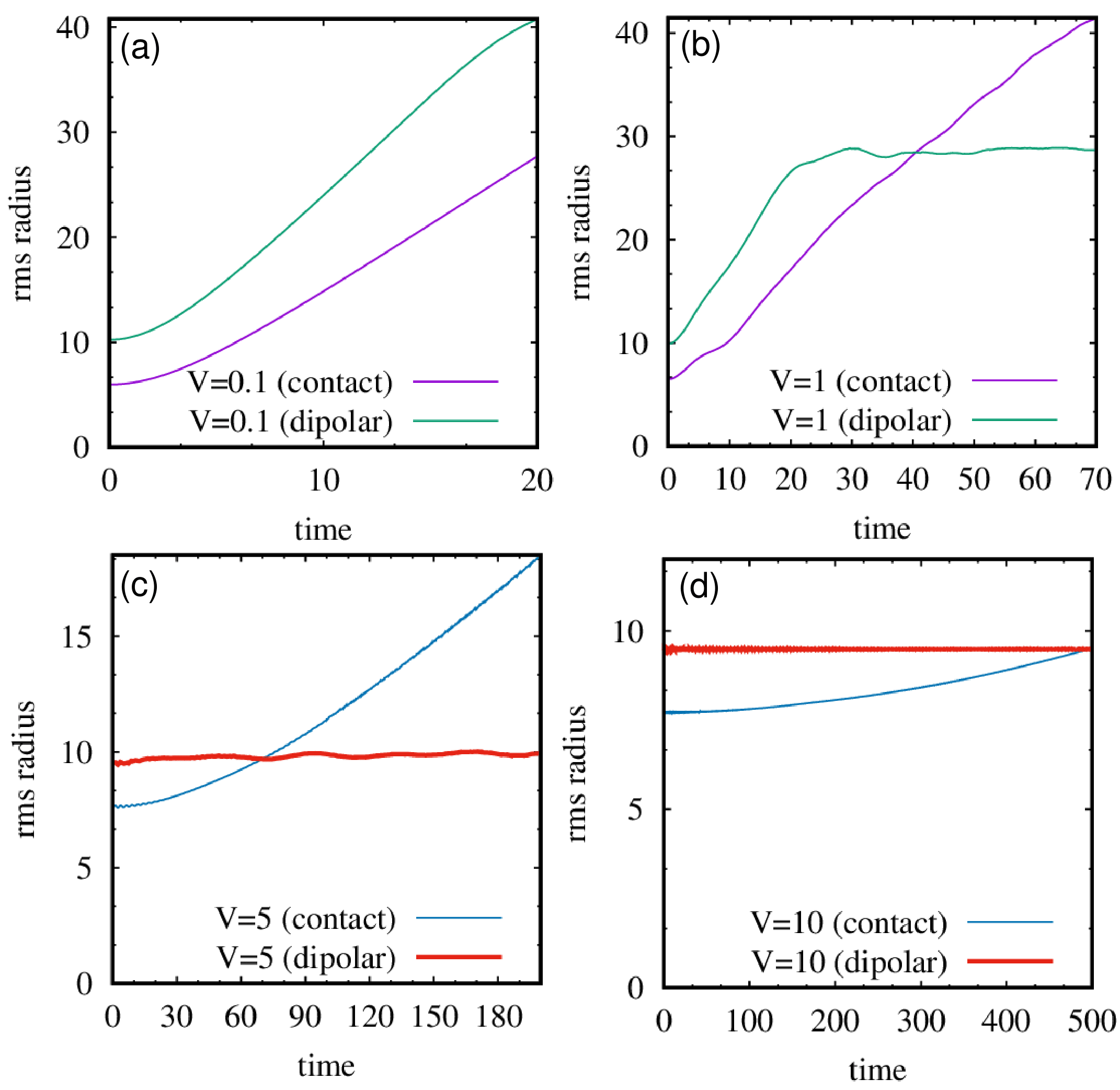}
\caption{Time evolution of the root mean square (rms) radius of the bosonic cloud for different lattice depths. Simulation is done with $N=4$ bosons in $S=21$ lattice sites and $M=15$. Contact interaction strength is fixed to $\lambda=25$ and dipolar interaction strength is kept fixed to $g_d=25$. (a) Expansion in a very shallow lattice ($V=0.1 E_r$);  both for contact as well as dipolar interaction the cloud expands, the dipolar bosons expand fast. (b) Expansion in an intermediate lattice ($V=1.0 E_r$); for contact interaction, the expansion is ballistic. For dipolar interaction, there is an initial expansion followed by the saturation of the cloud's radius which results to suppressed tunneling. (c) Expansion in a stronger lattice ($V=5.0 E_r$); in the case of contact interaction, the bosonic cloud expands over time. However, for dipolar interaction the localization happens, the corresponding rms radius remains constant. 
(d) Expansion in a deep lattice ($V=10.0$); for contact interaction the cloud remains arrested for some time and then expands. Whereas for dipolar interaction the transport remains arrested throughout the dynamics. All quantities are dimensionless.
}
\label{fig3}
\end{figure}

Here we prepare the initial state in the combination of harmonic trap ($\omega = 0.22$) and lattice with depth $V =0.1 E_r$. To induce dynamics, we suddenly switch off the harmonic trap and allow the bosons to tunnel in the outer wells.  The many-body expansion dynamics is studied for $M=15$ orbitals. On increasing  the number of orbitals to $M=16$, the occupation in the highest natural orbital becomes insignificant, it assures the convergence in the dynamical evolution. The one-body density evolution is presented in the Fig.~\ref{fig2}. The potential is effectively a harmonic oscillator potential, the lattice is sub relevant and the initial state is a fermionized phase for contact interaction and a crystal phase for dipolar interaction~Fig.~\ref{fig1}. Sudden quench to harmonic trap frequency to zero makes the bosons trapless and allows ballistic expansion as shown in Fig~\ref{fig2}. However, both for the contact and dipolar interaction, density dynamics exhibit distinguished many-body features in the expansion dynamics. At $t=0$, for contact interaction, four bright red spots signify four strongly interacting bosons. On switching off the trap they propagate as four bright jets till time $t=5.0$, the two inner jets are brighter than the two outer jets which agree well with the initial state density presented in Fig.~\ref{fig1}. After time $t=5.0$, the structure melts and the many-body features disappear, the cloud expands as a whole. In contrast, for the dipolar interaction, the jets are very bright and well separated as shown in lower panel of Fig.~\ref{fig2}. The distinct many-body feature is maintained in the expansion of crystallized bosons until they hit the wall, only with time of propagation the intensity of the jets is reduced. The diverging nature in the expansion dynamics of dipolar bosons signify the unbounded energy for dipolar interaction, in contrast, the expansion dynamics for contact interaction is bounded. The corresponding root mean square radius calculated from  $\sqrt{\int{x^{2}\rho(x)dx}}$ determines the average size of the expanding cloud and is plotted in Fig.~\ref{fig3}(a). Both exhibit ballistic expansion from time $t=5.0$, the initial expansion (just after release of the trap) has some nonlinear behaviour which looks the initial expansion is diffusive like. The cloud radius for dipolar bosons is more as expected. 

\subsection{Expansion dynamics in intermediate lattice: $V=1.0 E_r$}

\begin{figure}
\centering
\includegraphics[scale=0.25, angle=-90]{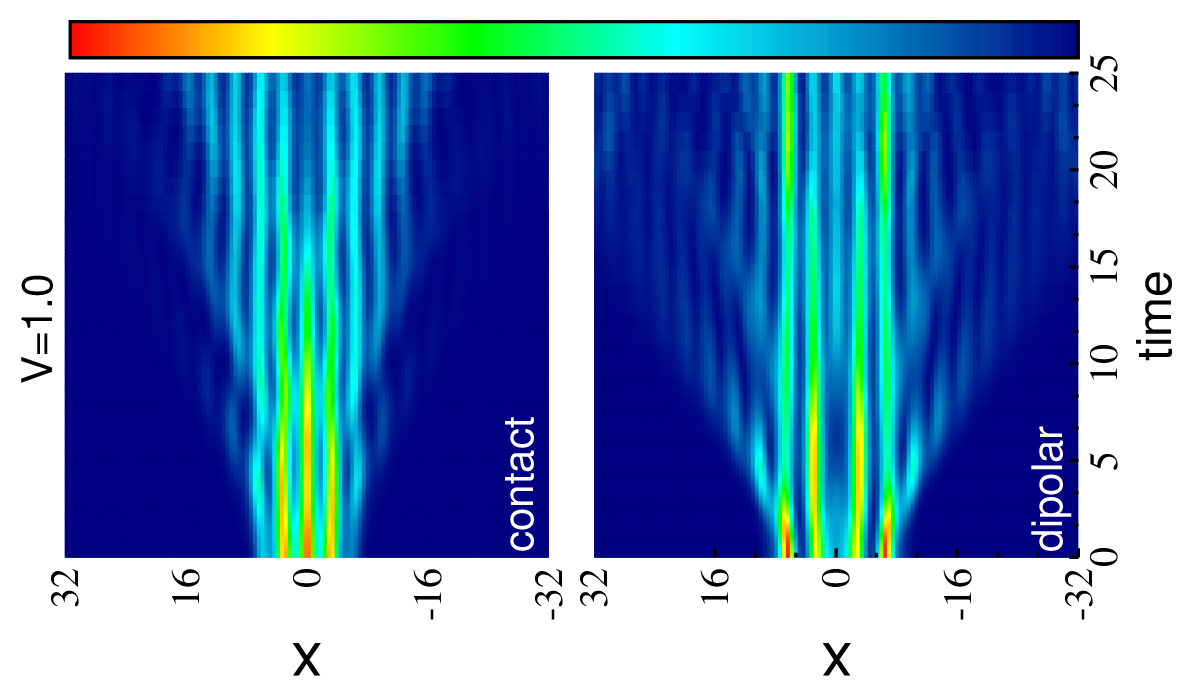}
\caption{ Density dynamics on sudden removal of trap in the Hamiltonian (Eq.(5)), for $N=4$ bosons, orbital $M=15$, lattice sites $S=21$, lattice depth $V=1.0 E_r$. The upper panel depicts post quench density dynamics for contact interaction ($\lambda=25$), while the lower panel shows the same for dipolar interaction ($g_d=25$). All quantities are dimensionless. See the text for discussion.}
\label{fig4}
\end{figure}
Expansion dynamics presented in  Fig.~\ref{fig4} exhibit bimodal structure both for the contact and dipolar interaction. The central cloud expansion which is coherent is surrounded by incoherent halo cloud in both cases. For contact interaction, the high density central cloud consists of three inner peaks (as shown in Fig.~\ref{fig1}), expands slowly. Whereas the outer low density halo cloud consists of two weak outer peaks (as shown in Fig.~\ref{fig1}) tunnelled out to the neighbouring lattice sites ballistically. The corresponding cloud radius as shown in Fig.~\ref{fig3}(b) also exhibit ballistic expansion, the cloud radius almost linearly grows up with time.  
The expansion dynamics of  dipolar bosons is distinctly different in time scale as shown in the lower panel of Fig.~\ref{fig4}. As the initial state as shown in Fig.~\ref{fig1} consists of four peaks almost of same intensity, the expansion dynamics is basically controlled by the expansion of central cloud surrounded by extremely fade halo structure. We observe expansion up to time $t=25$ as shown in Fig.~\ref{fig3}(b). At the very initial releasing point the expansion exhibits diffusive like for very short time and then becomes ballistic and remains till time $t=25$. However at longer time, the expansion speed is reduced and finally the transport is suppressed. This happens due to large mismatch in the  energy in neighbouring sites. The interaction energy costs tunneling and when there is a significant mismatch in energy in the adjacent lattice sites, tunneling of the bosons from central cloud is inhibited. Due to long range correlation in dipolar interaction, the bosons experience more mismatched correlation and favor to stay in the central lattice rather than tunneling. The localization also happens for contact interaction but at a much longer scale not shown in Fig.~\ref{fig4}. 

\subsection{Expansion dynamics in stronger lattice: $V=5.0 E_r$}

\begin{figure}
\centering
\includegraphics[scale=0.25, angle=-90]{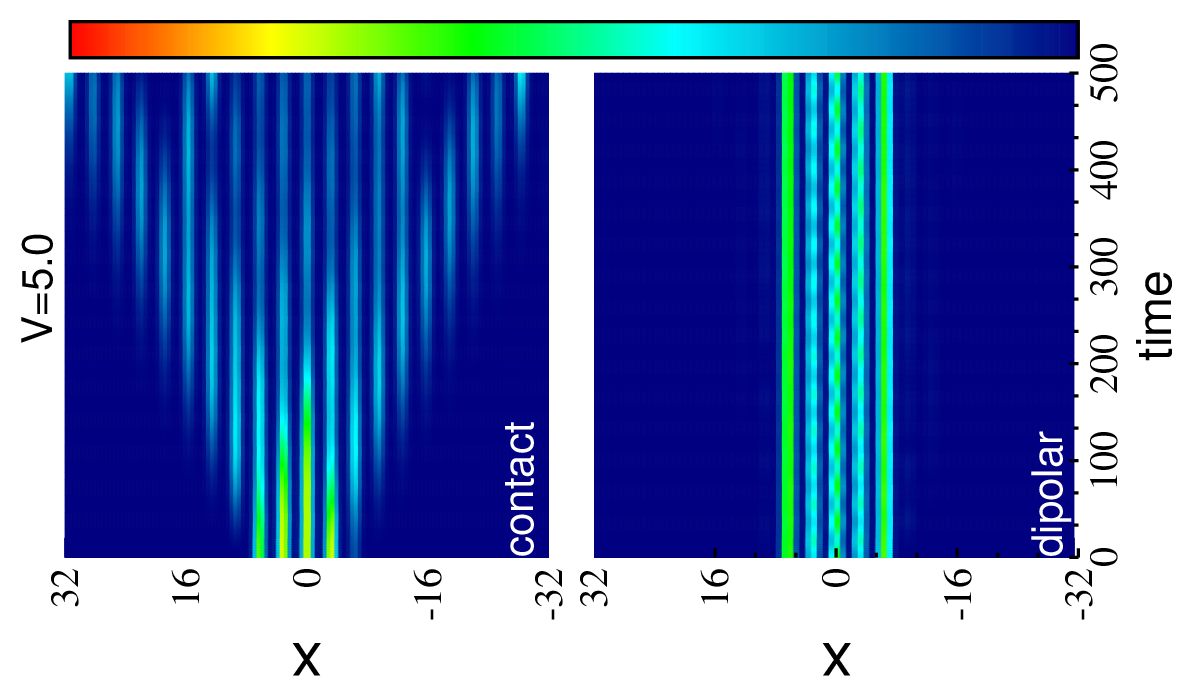}
\caption{ Density dynamics on sudden removal of trap in the Hamiltonian (Eq.(5)), for $N=4$ bosons, orbital $M=15$, lattice sites $S=21$, lattice depth $V=5.0 E_r$. The upper panel depicts post quench density dynamics for contact interaction ($\lambda=25$), while the lower panel shows the same for dipolar interaction ($g_d=25$). All quantities are dimensionless. See the text for discussion}
\label{fig5}
\end{figure}
The expansion dynamics of strongly interacting bosons in a lattice with depth $V=5.0 E_r$ is presented in Fig.~\ref{fig5}. For this choice of lattice depth, the effective lattice becomes inhomogeneous and has consequence in configuring the initial states for contact interaction and dipolar interaction as shown in Fig.~\ref{fig1}. We also observe distinct difference in the expansion dynamics due to dipolar interaction. For contact interaction, the central cloud consists of four bright jets propagate uniformly  till time $t=30$. Up to this time tunneling does not happen in the outer lattice, the corresponding rms radius remains constant as shown in Fig.~\ref{fig3}(c).
Then the central cloud disappears and quick tunneling happens in the outer lattice sites which is exhibited by the halo structure which further expands uniformly as a  bright wave front. The corresponding rms radius also uniformly increases. Thus we find two independent modes in the expansion dynamics. Initially it is contributed by the high density central cloud and later by low density outer halo.  In contrast, for dipolar bosons, the four high intensity peaks in the initial state remains like this all throughout the dynamics  as shown in the lower panel of Fig.~\ref{fig5}. The long range correlation causes mismatch in the energy between the adjacent sites as argued before, however the stronger lattice depth also inhibits the propagation which causes localization even in the long time dynamics. The corresponding rms radius is also constant as shown in Fig.~\ref{fig3}(c).

\subsection{Expansion dynamics in very strong lattice: $V=10.0 E_r$}
\begin{figure}
\centering
\includegraphics[scale=0.25, angle=-90]{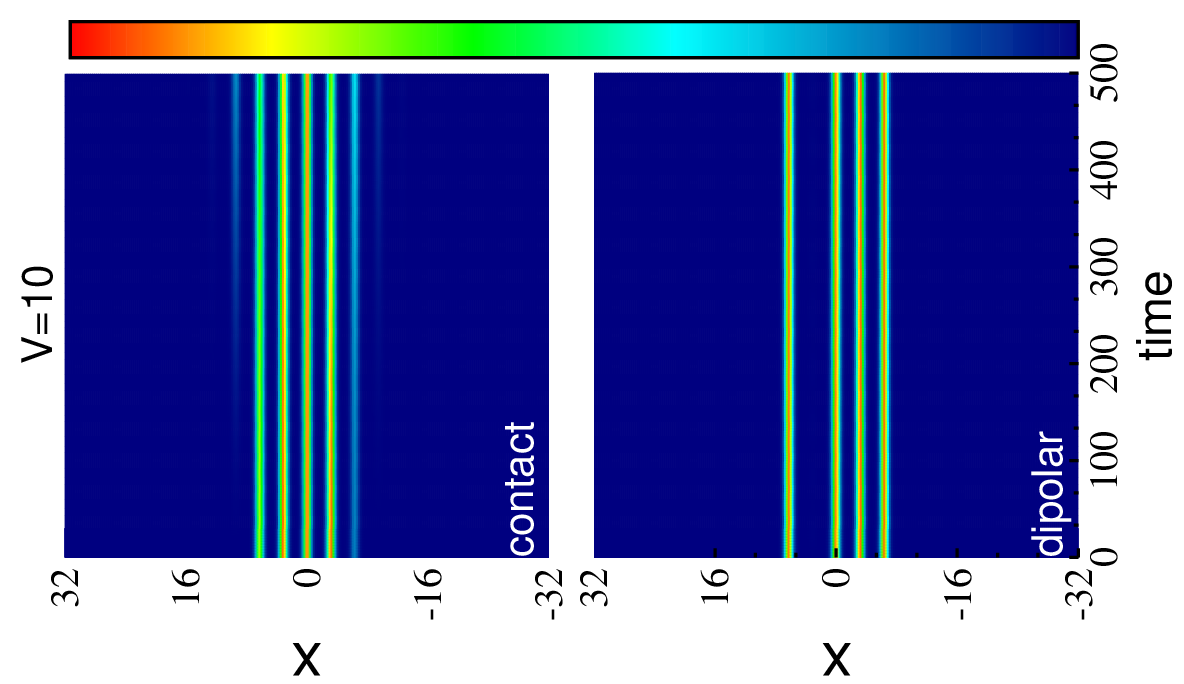}
\caption{ Density dynamics on sudden removal of trap in the Hamiltonian (Eq.(5)), for $N=4$ bosons, orbital $M=15$, lattice sites $S=21$, lattice depth $V=10.0E_r$. The upper panel depicts post quench density dynamics for contact interaction ($\lambda=25$), while the lower panel shows the same for dipolar interaction ($g_d=25$). All quantities are dimensionless. See the text for discussion.}
\label{fig6}
\end{figure}

The dynamics is studied in a deep lattice of lattice depth $V=10.0 E_r$. The corresponding density dynamics is presented in Fig~\ref{fig6}. Arrested transport is expected for both types of interaction as strong lattice plays the dominating role and will inhibit the tunneling. However, for contact interaction bosons remain localized up to certain time and then the cloud starts to tunnel in the adjacent wells only.  Where as, for dipolar interaction we observe complete localization. The highly correlated dipolar bosons need to pay much larger interaction energy for tunneling. The cost of larger interaction energy is justified as dipolar bosons feel more mismatch in the adjacent site correlation due to asymmetry in the density compared to the non-dipolar boosns. The same dynamical features are concluded from  Fig~\ref{fig3}(d) where cloud radius is initially constant and then slowly increases for contact interaction whereas for dipolar interaction the rms radius remains constant throughout the entire dynamics.

\section{Conclusion}\label{conc}
We have investigated the 
expansion dynamics of strongly correlated dipolar bosons in crystallized limit in 1D homogeneous lattices.  Our protocol consists of preparing initial states in a different lattice depth superimposed with a harmonic trap and then suddenly switching off the trap. The dynamics is studied by key measures of one-body density dynamics, rms radius of the expanding cloud, one-body and two-body correlation dynamics. All results are compared with the expansion dynamics for the strongly interacting bosons in fermionized limit. We maintain the same quench protocol for fermionized and crystallized boson expansion dynamics. We observe very interesting and distinguished many-body features in the expansion dynamics for all initial setups. For very weak lattice depths, when the harmonic potential trap plays the dominating role to configure the initial set up, both the crystallized and fermionized bosons exhibit ballistic expansion. However the ballistic expansion in two cases can be distinguished utilizing the many-body features. The long range tail in dipolar interaction allows the crystallized bosons to carry on the independent jet like structure until it hits the boundary of the lattice.  Whereas the fermionized bosons quickly loose the many body features, and the cloud expands ballistically.  On the other habd, very strong lattice inhibits the transport as expected.  The most intriguing many-body features are observed for intermediate lattice; we observe strong interplay between lattice depth, strong interaction and long range correlation. The cloud expansion is separated into central cloud expansion with distinct many body features and a halo low density cloud expansion without many body features. The expansion is bimodal, however the many body features are different for the crystallized bosons from those of fermionized bosons. With increase in lattice depth, the dipolar bosons exhibit localization as they are completely isolated due to long range interaction and the correlation in the adjacent wells are strongly mismatched. Thus tunneling is not favored at the cost of large interaction energy. Instead of paying large cost, dipolar bosons like to stay in the well which results to arrested transport. Thus our present study highlights how to manipulate the initial state configuration and interaction to control the tunneling dynamics. 

The kind of initial states prescribed in this manuscript should be established in the current experimental setups and to study the many-body correlation in the strongly interacting limit. In the future, the similar approach can be made for localized fermions both for the attractive and repulsive interaction. Our analysis can be extended to study the other crystallization phenomena in higher
dimensions or for other kind of long-range interactions, e.g. mediated by cavities~\cite{Molignini:2022,Lukas:2023}. We expect that several of the qualitative results discussed in this paper hold also for larger values of $N$, and the exploration of regimes with large $N$ is an interesting subject of future investigation. Finally, we observe that the expansion dynamics may be strongly influenced in the presence of disordered potential, which is an interesting topic for future 
work.

\begin{acknowledgements}
 We would like to thank N. Defenu, T. Macri', and A. Smerzi for useful discussions. BC acknowledges the financial support  and hospitality of ICTP.
\end{acknowledgements}
\appendix
\section{ The Multiconfigurational Time dependent Hartree method for bosons}
The ansatz for the many-body wave function is the linear combination of time dependent permanents
\begin{equation}
\vert \Psi(t)\rangle = \sum_{\bar{n}}^{} C_{\bar{n}}(t)\vert \bar{n};t\rangle,
\label{many_body_wf}
\end{equation}
The vector $\vec{n} = (n_1,n_2, \dots ,n_M)$ represents the occupation of the orbitals, with the constraint that $n_1 + n_2 + \dots +n_M = N$, which ensures the preservation of the total number of particles. 
Distributing $N$ bosons over $M$ time dependent orbitals, the number of permanents become  $ \left(\begin{array}{c} N+M-1 \\ N \end{array}\right)$.
The permanents are constructed over $M$ time-dependent single-particle wave functions, called orbitals, as 
\begin{equation}
\vert \bar{n};t\rangle = \prod^M_{k=1}\left[ \frac{(\hat{b}_k^\dagger(t))^{n_k}}{\sqrt{n_k!}}\right] |0\rangle 
\label{many_body_wf_2}
\end{equation}

Where $|0\rangle$ is the vacuum state and $\hat{b}_k^\dagger(t)$ denotes the time-dependent operator that creates one boson in the $k$-th working orbital $\psi_k(x)$, \textit{i.e.}:
\begin{eqnarray}
	\hat{b}_k^\dagger(t)&=&\int \mathrm{d}x \: \psi^*_k(x;t)\hat{\Psi}^\dagger(x;t) \:  \\
	\hat{\Psi}^\dagger(x;t)&=&\sum_{k=1}^M \hat{b}^\dagger_k(t)\psi_k(x;t). \label{eq:def_psi}
\end{eqnarray}
The accuracy of the algorithm depends on the number of orbitals $M$.
$M=1$ corresponds to the mean field Gross-Pitaevskii equation.
If $M \rightarrow \infty$, the wave function becomes exact, with the set $ \vert n_1,n_2, \dots ,n_M \rangle$ spanning the complete Hilbert space for $N$ particles. However, due to computational limitations, the number of orbitals is restricted to a desired value to ensure proper convergence in the measured quantities. It is important to note that both the expansion coefficients $\left \{C_{\bar{n}}(t)\right\}$ and the working orbitals $\psi_i(x,t)$ that are used to construct the permanents $\vert \bar{n};t\rangle$ are fully variationally optimised and time-dependent quantities~\cite{TDVM81,variational1,variational3,variational4}. The time-dependent optimization of the orbitals and expansion coefficients ensure that the MCTDHB method can accurately describe the dynamics of strongly interacting bosons. As the permanents are time-dependent, a given degree of accuracy is achieved with shorter time compared to a time-independent basis.
The time-adaptive many-body basis set employed in MCTDHB allows for the dynamic tracking of correlations that arise from inter-particle interactions.

\section{Measures of correlation}
\begin{figure}
\centering
\includegraphics[scale=0.18, angle=0]{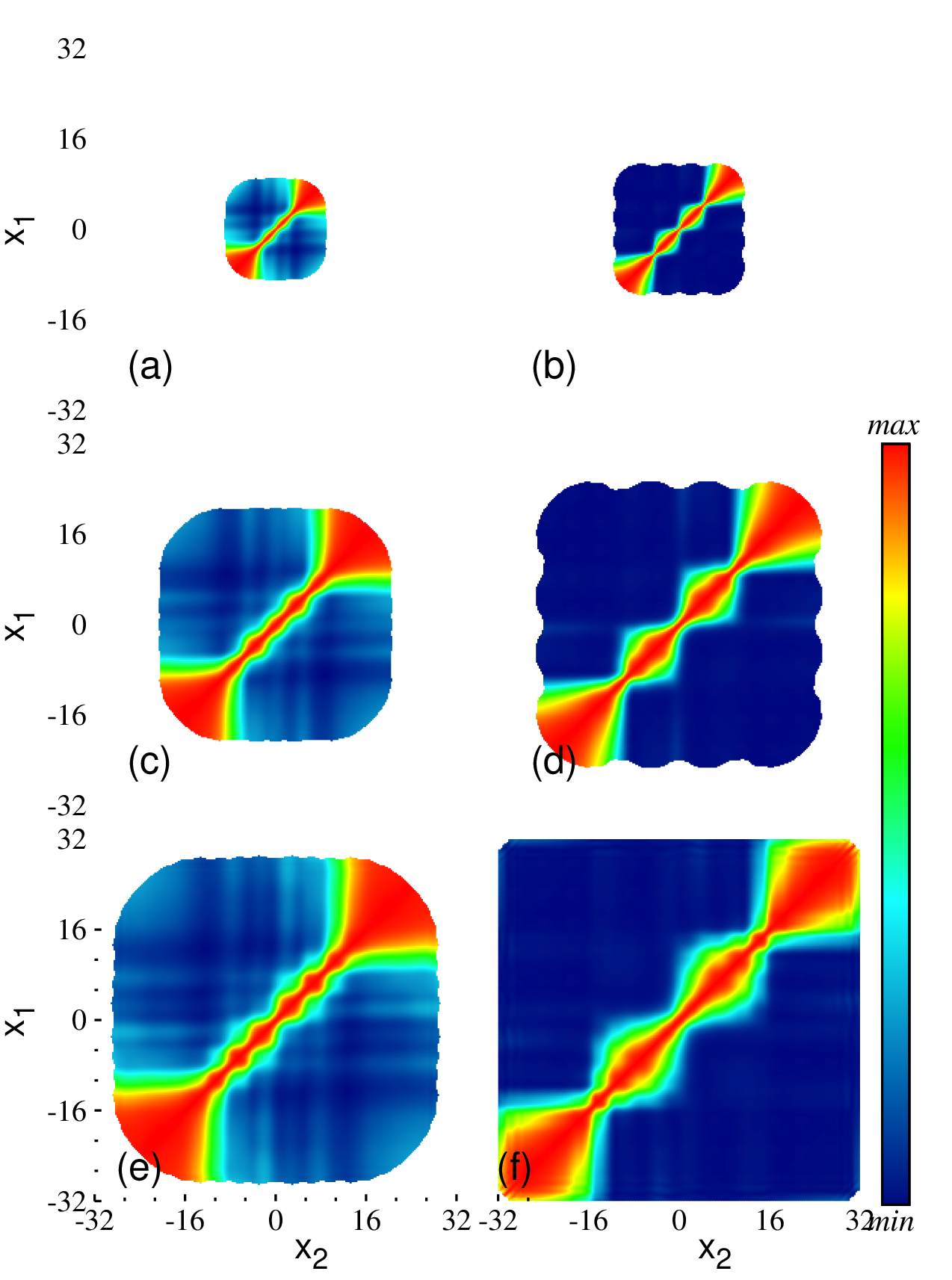}
\caption{Dynamics of the one-body Glauber correlation function $g^{(1)}(x_1,x_2)$ for lattice depth $V=0.1 E_r$. Figs. (a), (c), and (e) are for contact interaction with $\lambda=25$, while Figs. (b), (d), and (f) are for dipolar interaction with $g_d=25$. The pre-quench and post quench parameters remain same as in the expansion dynamics. The plots show the correlation functions at three different times: $t=0.0$, $t=10.0$, and $t=15.0$ (top to bottom).}
\label{fig7}
\end{figure}

\begin{figure}
\centering
\includegraphics[scale=0.18, angle=0]{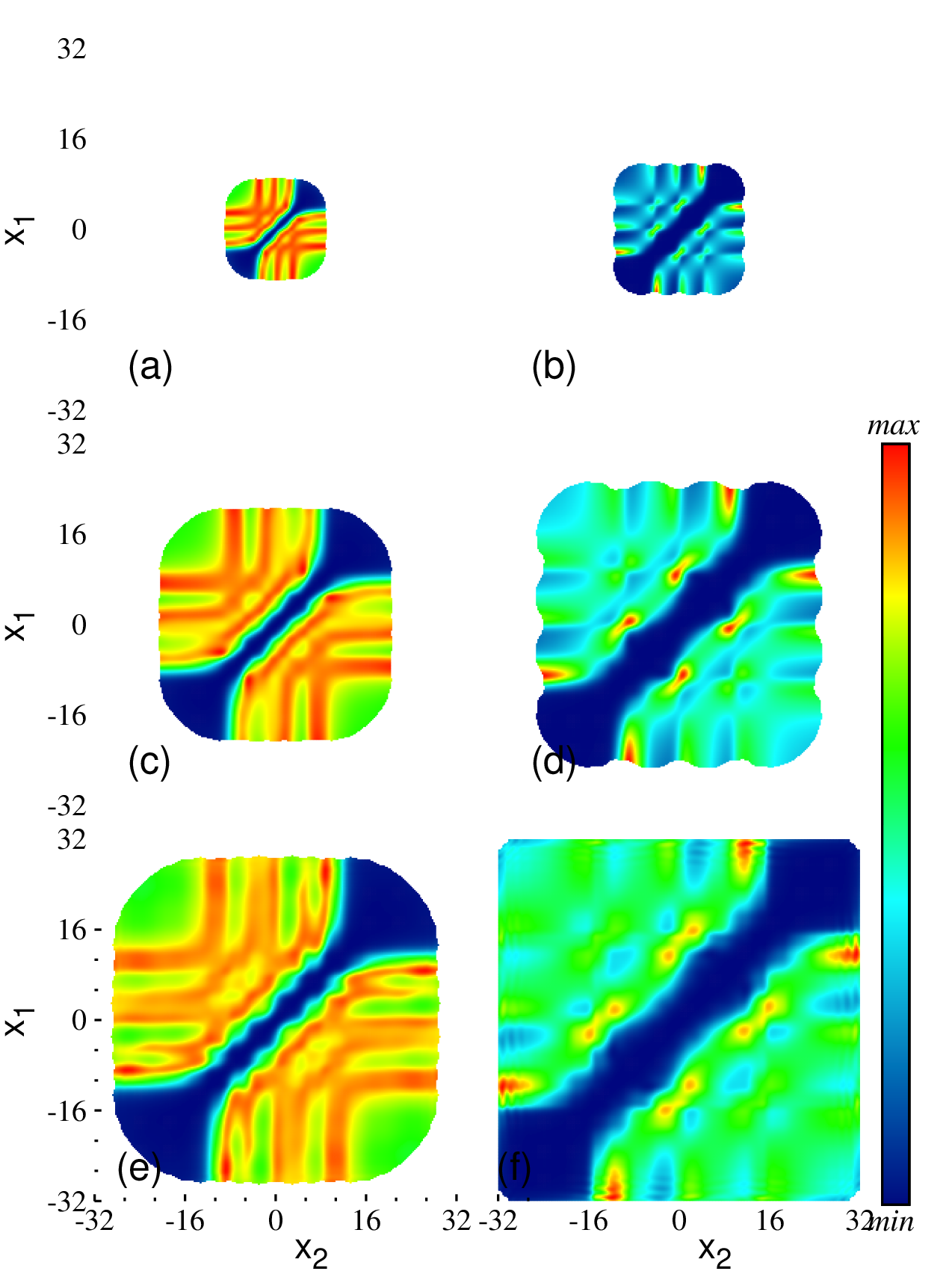}
\caption{Dynamics of the two-body Glauber correlation function  $g^{(2)}(x_1,x_2)$ for lattice depth $V=0.1E_r$. Figs. (a), (c), and (e) are for contact interaction with $\lambda=25$, while Figs. (b), (d), and (f) are for dipolar interaction with $g_d=25$.  The pre-quench and post quench parameters remain same as in the expansion dynamics. The plots show the correlation function at three different times: $t=0.0$, $t=10.0$, and $t=15.0$ (top to bottom).}
\label{fig8}
\end{figure}

In this appendix, we present the emergent correlation properties by the measures of one-body and two-body Glauber correlation functions as defined in Eqs.(2) to (4). The parameters for quench dynamics remain the same as described in the expansion dynamics. The results for very weak lattice $(V=0.1E_r)$ are presented in Figs.7 and 8 for some selected time points $t=0.0$, $10.0$ and $15.0$. At $t=0$ (pre-quench state), the diagonal of the first order correlation for contact interaction (Fig.~\ref{fig7}(a)) shows four non isolated bright regions where $|g^{(1)}|^{2}$
$\simeq$ $1$. This describes correlation between four fermionized bosons trapped in the HO potential. The corresponding for dipolar bosons (Fig.~\ref{fig7}(b)), shows four completely separated bright lobes. This describes correlation between four crystallized bosons, it is specifically to be  noted that bright lobes are completely isolated due to long range interaction. The corresponding $|g^{(2)}|$ in Figs.~\ref{fig8}(a) (contact interaction) and (b) (dipolar interaction) shows four extinguished lobes, which are termed as 'correlation holes'. It signifies that the probability of finding two bosons in the same place $(x_1=x_2)$ is zero. On sudden removal of the trap, the correlations, both one- and two-body exhibit expansion for contact as well as dipolar interaction. The structure of diagonal correlation in $g^{(1)}$ is maintained during expansion, whereas the width of the correlation hole in $g^{(2)}$ also expands maintaining the distinguished features.

\begin{figure}
\centering
\includegraphics[scale=0.18, angle=-90]{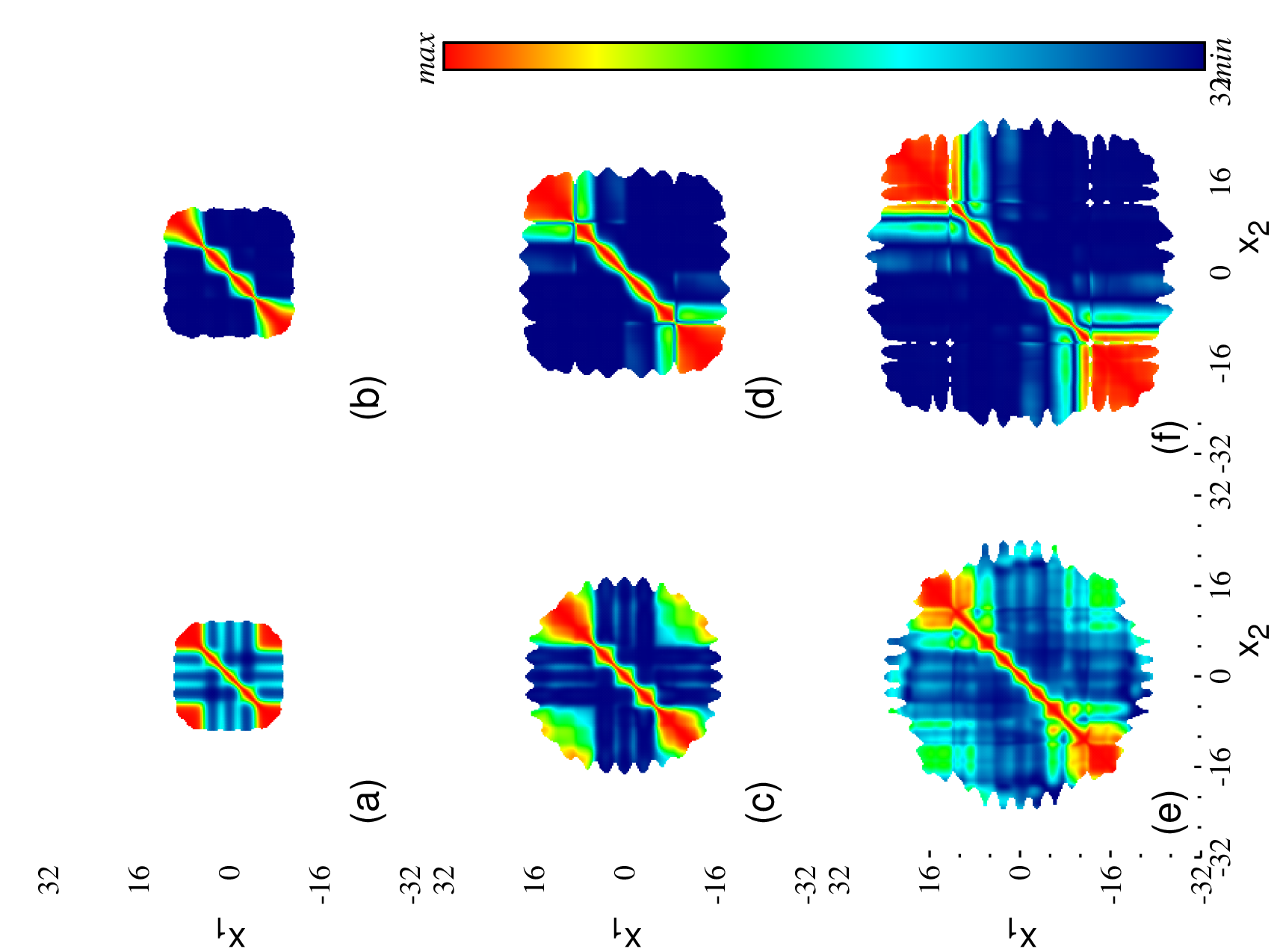}
\caption{Dynamics of the one-body Glauber correlation function $g^{(1)}(x_1,x_2)$ for lattice depth $V=1.0 E_r$. Figs. (a), (c), and (e) are for contact interaction with $\lambda=25$, while Figs. (b), (d), and (f) are for dipolar interaction with $g_d=25$. The pre-quench and post quench parameters remain same as in the expansion dynamics. The plots show the correlation functions at three different times: $t=0.0$, $t=5.0$, and $t=10.0$ (top to bottom).}
\label{fig9}
\end{figure}

\begin{figure}
\centering
\includegraphics[scale=0.18, angle=-90]{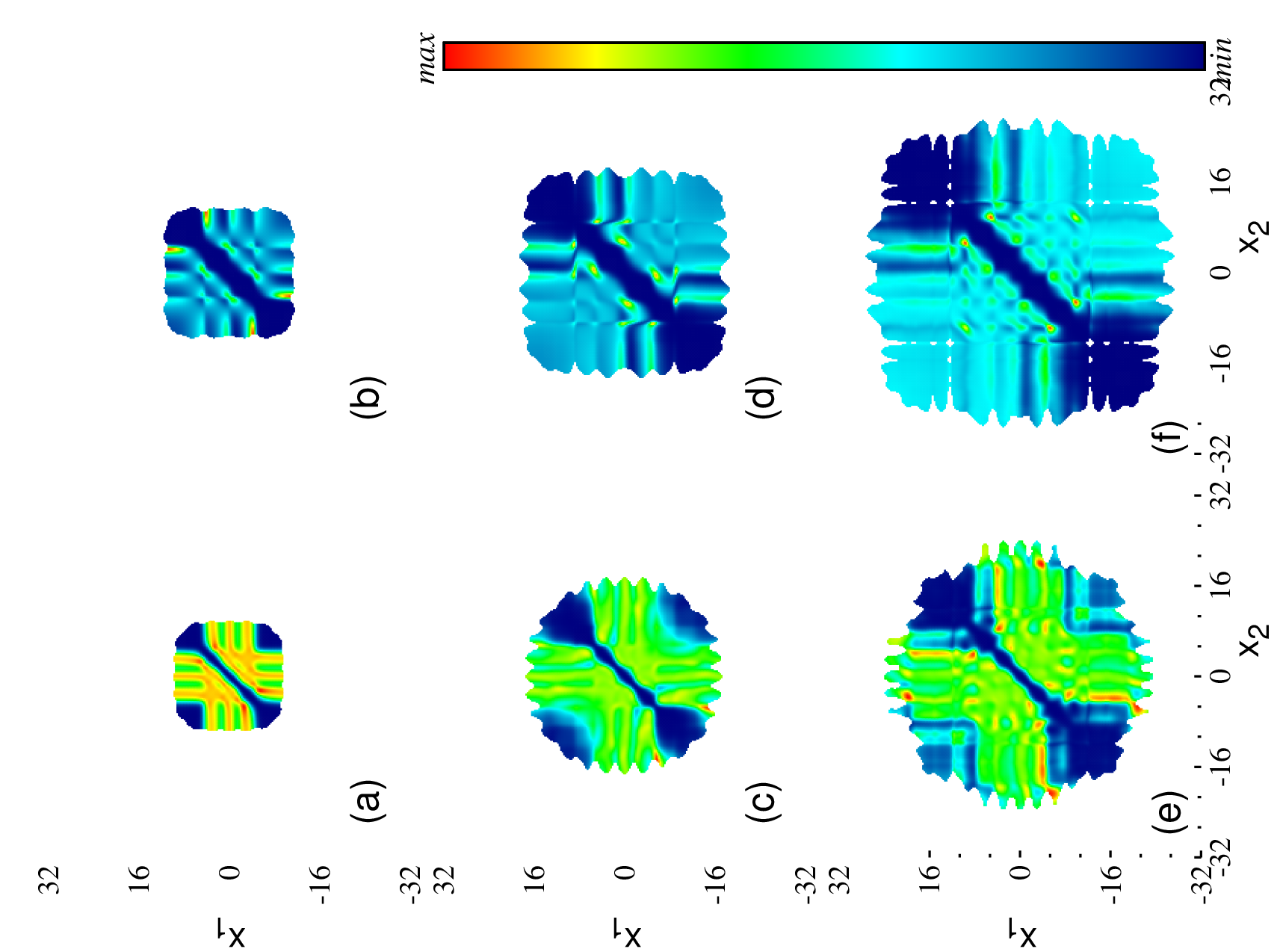}
\caption{Dynamics of the two-body Glauber correlation function  $g^{(2)}(x_1,x_2)$ for lattice depth $V=1.0 E_r$. Figs. (a), (c), and (e) are for contact interaction with $\lambda=25$, while Figs. (b), (d), and (f) are for dipolar interaction with $g_d=25$.  The pre-quench and post quench parameters remain same as in the expansion dynamics. The plots show the correlation function at three different times: $t=0.0$, $t=5.0$, and $t=10.0$ (top to bottom).}
\label{fig10}
\end{figure}

 Fig.~\ref{fig9} displays the spatial first-order correlation function for a lattice depth $V=1.0 E_r$. Figs.~\ref{fig9}(a), (c), and (e) depict the correlation function for contact interaction with $\lambda=25$, while Figs.~\ref{fig9}(b), (d), and (f) show the correlation function for dipolar interaction with $g_d=25$. The plots are presented at three different times: $t=0.0$, $t=5.0$, and $t=10.0$.
At $t=0$, the diagonal of the first-order correlation function exhibits coherent regions both for contact Fig.~\ref{fig9}(a) and dipolar interaction Fig.~\ref{fig9}(b). In the case of contact interaction, two small red patches at top left corner and down right corner appear. It signifies the two small peaks as observed in the initial state (Fig.~\ref{fig1}). 
Fig.~\ref{fig9}(b) signifies that the dipolar bosons are initially more isolated than the case of contact interaction. As time progresses, the correlations expand and occupy the allotted space and in consequence the off diagonal coherence fades up gradually for the case of contact interaction; Fig.~\ref{fig9}(c) and Fig.~\ref{fig9}(e). In comparison, Fig.~\ref{fig9}(d) and (f)  show that the correlation for dipolar bosons  expands very fast and without loss of off diagonal correlation. Thus the correlation quickly reaches the boundary and the further expansion in correlation dynamics stops. Thus at much longer time (not shown here), while the correlation expands for contact interaction, that for dipolar interaction reaches its saturation value. To determine the degree of spatial second-order coherence, we study the normalized two-body correlation function $g^{(2)}(x_1,x_2)$ and present for selected evolution times as before in the Fig.\ref{fig10}. Figs.\ref{fig10}(a), (c), and (e) correspond to contact interactions, while Figs.~\ref{fig10}(b), (d), and (f) correspond to dipolar interaction. At the initial time $t=0.0$, the emergence of a correlation hole along the diagonal of the normalized two-body correlation function $g^{(2)} (x_1, x_2=x_1)$ with $g^{(2)} (x_1, x_2=x_1) \rightarrow 0$ is a characteristic feature of the many-body system. This indicates that the bosons tend to maximize their spatial separation in the presence of strong repulsive interactions. The width of the correlation hole is broader for the dipolar interaction case.
 As the trap is switched off, the system expands and the two-body correlations spread. During expansion, the dynamics of contact and dipolar interactions can be distinguished by a single feature: the dipolar interaction leads to a larger width of the correlation hole along the diagonal. Unlike the case of one-body correlation, here we do not find any signature of the loss of off diagonal correlation.   
\\


\begin{figure}
\centering
\includegraphics[scale=0.18, angle=-90]{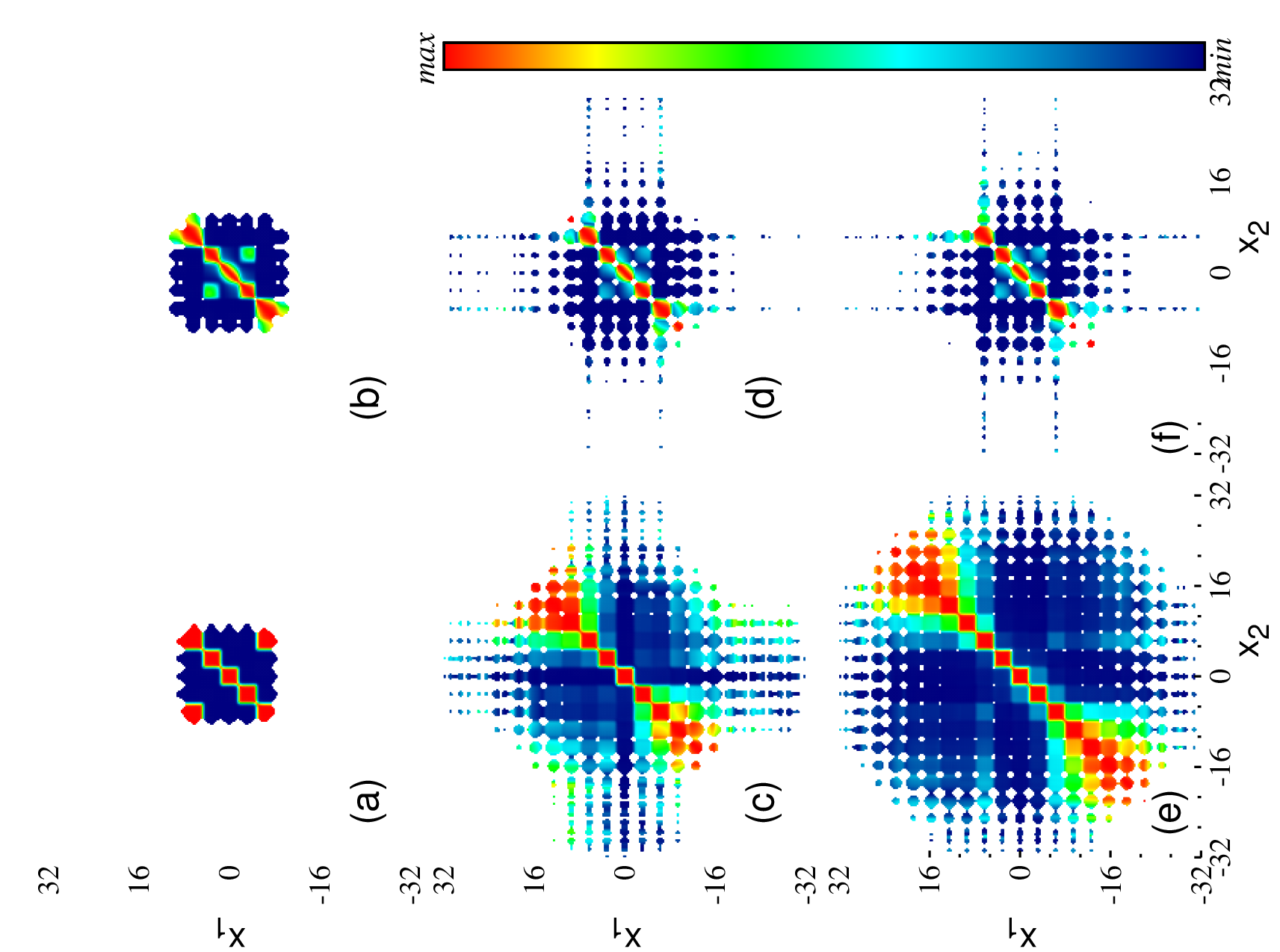}
\caption{Dynamics of the one-body Glauber correlation function $g^{(1)}(x_1,x_2)$ for lattice depth $V=5.0 E_r$. Figs. (a), (c), and (e) are for contact interaction with $\lambda=25$, while Figs. (b), (d), and (f) are for dipolar interaction with $g_d=25$. The pre-quench and post quench parameters remain same as in the expansion dynamics. The plots show the correlation functions at three different times: $t=0.0$, $t=50.0$, and $t=100.0$ (top to bottom).}
\label{fig11}
\end{figure}

In Fig.~\ref{fig11}, we present the one-body correlation for contact and dipolar interactions at different times. Figs.\ref{fig11}(a), (c), and (e) display the correlation function for contact interactions with $\lambda=25$, while Figs.\ref{fig11}(b), (d), and (f) depict the correlation function for dipolar interactions with $g_d=25$. The plots are shown at three different times: $t=0.0$, $t=50.0$, and $t=100.0$. At $t=0$, the normalized first-order correlation function for contact interaction (Fig.\ref{fig11}(a)) exhibits a structure along the diagonal which signify the distinct peaks in the initial state (Figs.\ref{fig1}) and $[g^{(1)}(x_{1},x_2)]^2 = 0$ on the off-diagonal. On expansion, diagonal correlation structure is completely maintained.  
 On the other hand, for dipolar interaction, at $t=0$, the diagonal structure is very prominent (Fig.\ref{fig11}(b)). On switching of the trap, the correlation does not expand  even for long time dynamics. During this 'arrested transport' regime, the one-body correlation along the diagonal is also maintained, without showing any expansion (Figs.\ref{fig11}((d) and (f)).

\begin{figure}
\centering
\includegraphics[scale=0.18, angle=-90]{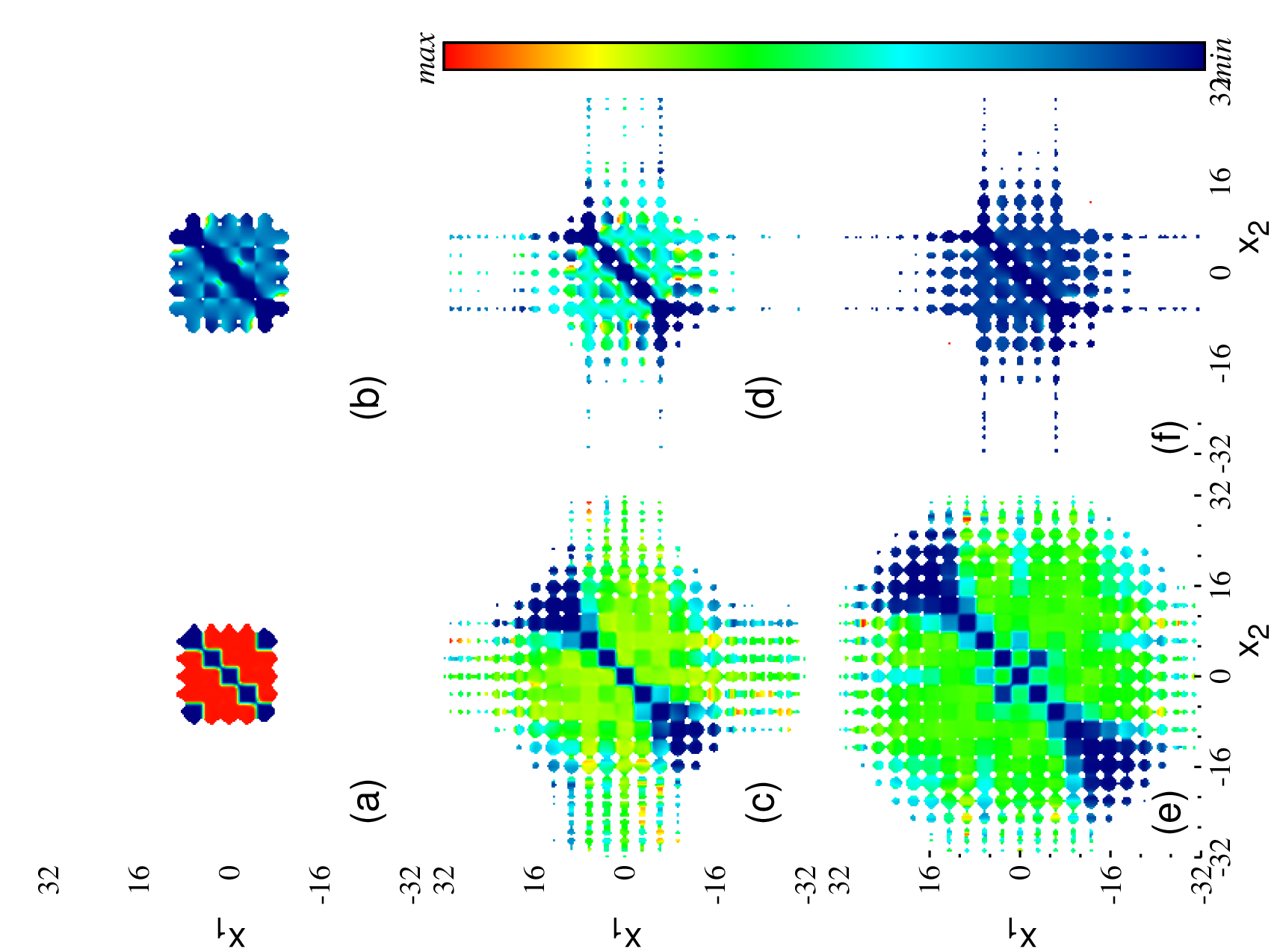}
\caption{Dynamics of the two-body Glauber correlation function  $g^{(2)}(x_1,x_2)$ for lattice depth $V=5.0 E_r$. Figs. (a), (c), and (e) are for contact interaction with $\lambda=25$, while Figs. (b), (d), and (f) are for dipolar interaction with $g_d=25$.  The prequench and post quench parameters remain same as in the expansion dynamics. The plots show the correlation function at three different times: $t=0.0$, $t=50.0$, and $t=100.0$ (top to bottom).}
\label{fig12}
\end{figure}
To assess the extent of spatial second-order coherence, we present the normalized two-body correlation function $g^{(2)}(x_1,x_2)$ during the expansion in
(Fig.\ref{fig12}). Figs.\ref{fig12}(a), (c), and (e) represent the correlation function for contact interactions with $\lambda=25$, while Figs.~\ref{fig12}(b), (d), and (f) correspond to dipolar interactions with $g_d=25$. The plots are presented at three distinct time points: $t=0.0$, $t=50.0$, and $t=100.0$. For contact interaction (Fig.~\ref{fig12}(a)), coherence is maintained at the off-diagonal elements ($g^{(2)}(x_1,x_2) \approx 1$), indicating that there is second-order coherence between different wells. However, along the diagonal, the correlation function exhibits a vanishing behavior, known as the correlation hole. In contrast, for dipolar interaction (Fig.~\ref{fig12}(b)), the correlation hole is broader, and no distinct square-like structure is formed. This behavior arises due to the long-range nature of the interaction. During the expansion, the correlation function $g^{(2)}(x_1,x_2)$ expands in an organized manner maintaining its gross features for contact interaction. However, for dipolar interaction, the two-body correlation as a whole presents restricted but gradual loss of second-order coherence within and between wells.  

\begin{figure}
\centering
\includegraphics[scale=0.18, angle=0]{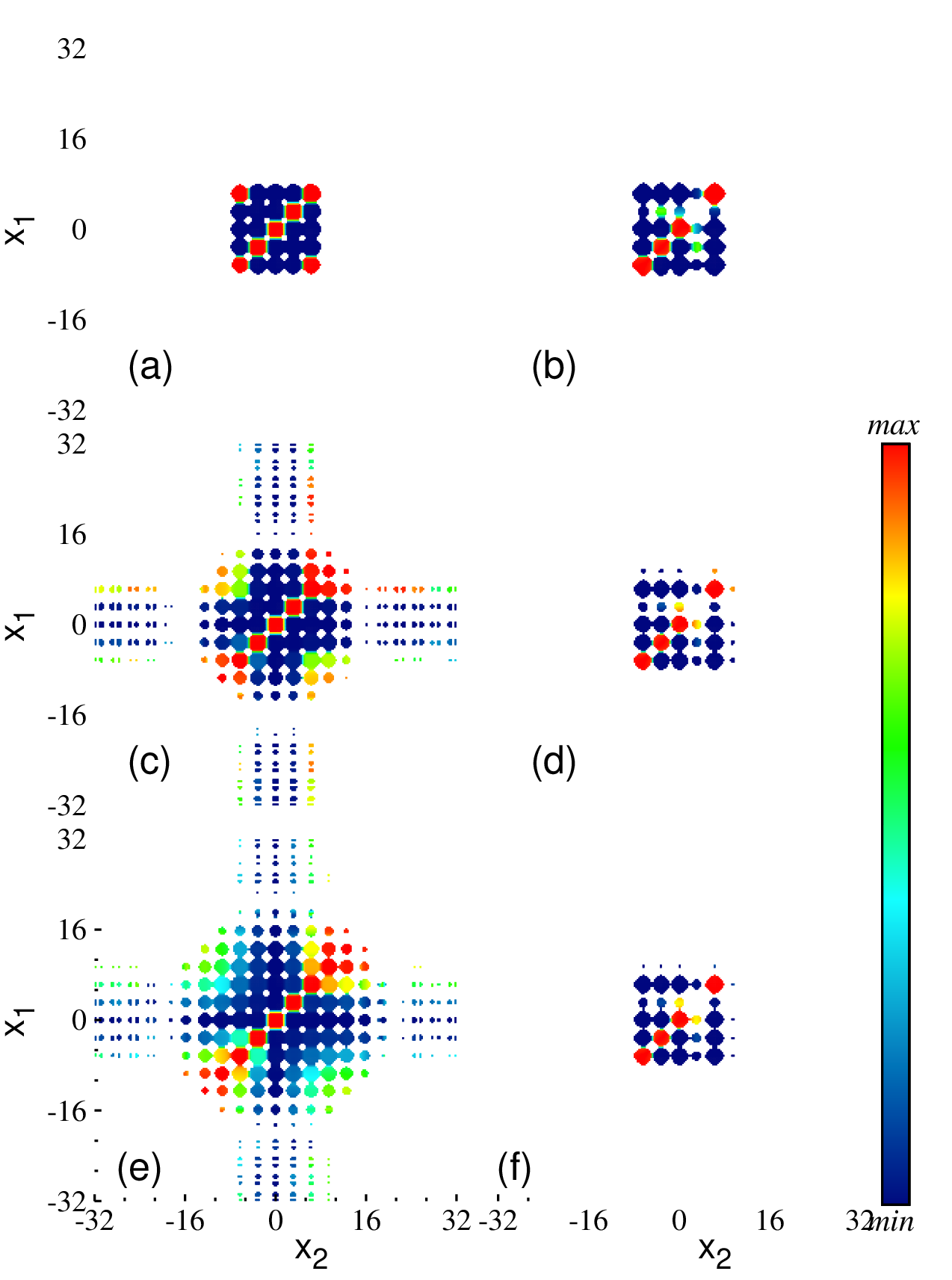}
\caption{Dynamics of the one-body Glauber correlation function $g^{(1)}(x_1,x_2)$ for lattice depth $V=10.0 E_r$. Figs. (a), (c), and (e) are for contact interaction with $\lambda=25$, while Figs. (b), (d), and (f) are for dipolar interaction with $g_d=25$. The prequench and post quench parameters remain same as in the expansion dynamics. The plots show the correlation functions at three different times: $t=0.0$, $t=200.0$, and $t=500.0$ (top to bottom).}
\label{fig13}
\end{figure}

\begin{figure}
\centering
\includegraphics[scale=0.18, angle=0]{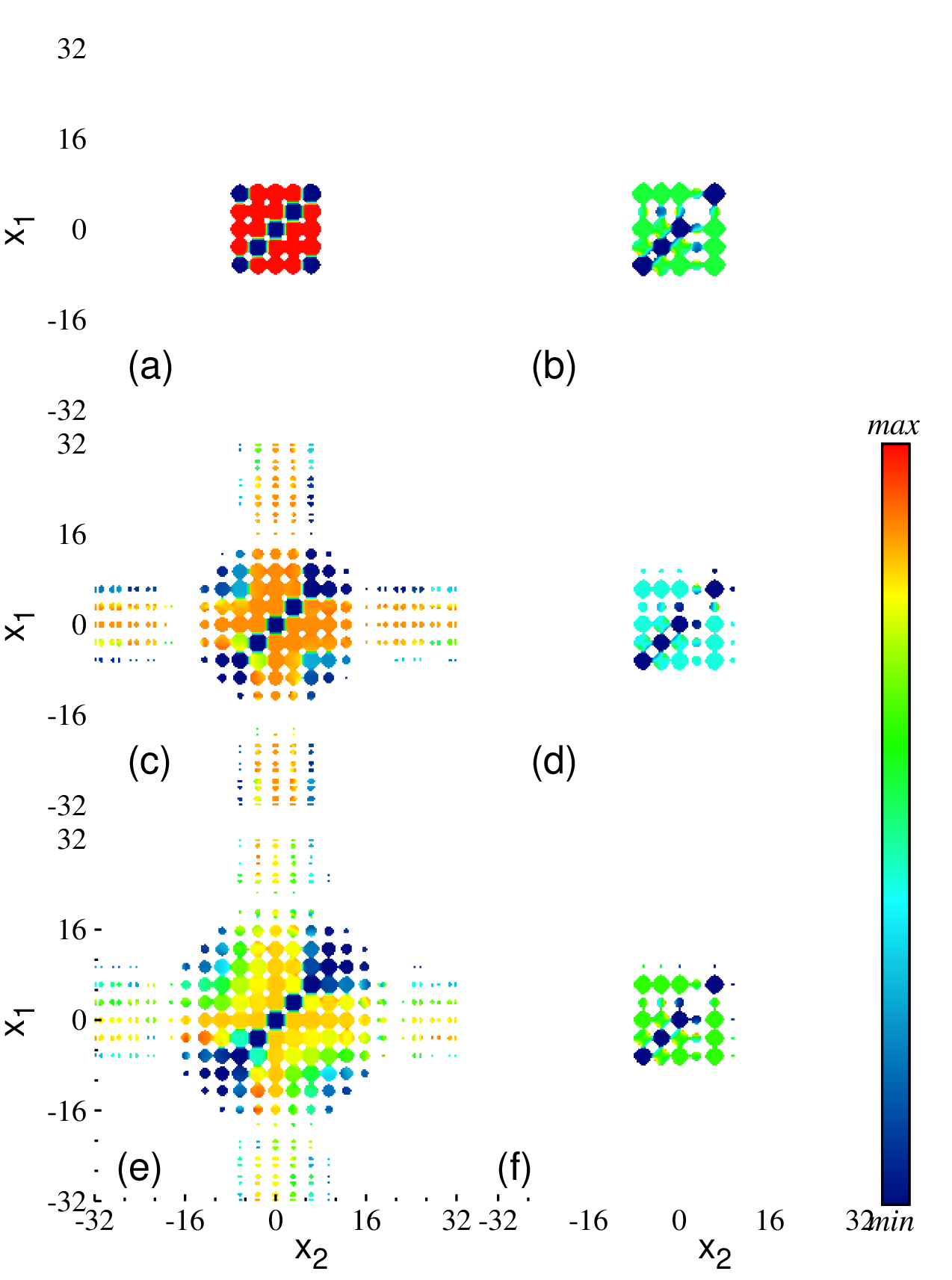}
\caption{Dynamics of the two-body Glauber correlation function  $g^{(2)}(x_1,x_2)$ for lattice depth $V=10.0 E_r$. Figs. (a), (c), and (e) are for contact interaction with $\lambda=25$, while Figs. (b), (d), and (f) are for dipolar interaction with $g_d=25$.  The pre-quench and post quench parameters remain same as in the expansion dynamics. The plots show the correlation function at three different times: $t=0.0$, $t=200.0$, and $t=500.0$ (top to bottom).}
\label{fig14}
\end{figure}

Fig.\ref{fig13} represents the one-body correlation in the deep lattice, $V=10.0E_r$ for selected times $t=0.0$, $200.0$ and $500.0$ to include the long time scenario. The diagonal structure in Figs.\ref{fig11}(a) (contact interaction) and Figs.\ref{fig11}(b) (dipolar interaction) characterize the initial state correlation (Figs.\ref{fig1}). On switching of the trap, the correlation initially arrested but then expands diffusively for contact interaction whereas the correlation for dipolar interaction is completely arrested even for very long time. The corresponding two-body correlation in Figs.\ref{fig14} also concludes the same physics.

\begin{thebibliography}{65}%
\makeatletter
\providecommand \@ifxundefined [1]{%
 \@ifx{#1\undefined}
}%
\providecommand \@ifnum [1]{%
 \ifnum #1\expandafter \@firstoftwo
 \else \expandafter \@secondoftwo
 \fi
}%
\providecommand \@ifx [1]{%
 \ifx #1\expandafter \@firstoftwo
 \else \expandafter \@secondoftwo
 \fi
}%
\providecommand \natexlab [1]{#1}%
\providecommand \enquote  [1]{``#1''}%
\providecommand \bibnamefont  [1]{#1}%
\providecommand \bibfnamefont [1]{#1}%
\providecommand \citenamefont [1]{#1}%
\providecommand \href@noop [0]{\@secondoftwo}%
\providecommand \href [0]{\begingroup \@sanitize@url \@href}%
\providecommand \@href[1]{\@@startlink{#1}\@@href}%
\providecommand \@@href[1]{\endgroup#1\@@endlink}%
\providecommand \@sanitize@url [0]{\catcode `\\12\catcode `\$12\catcode `\&12\catcode `\#12\catcode `\^12\catcode `\_12\catcode `\%12\relax}%
\providecommand \@@startlink[1]{}%
\providecommand \@@endlink[0]{}%
\providecommand \url  [0]{\begingroup\@sanitize@url \@url }%
\providecommand \@url [1]{\endgroup\@href {#1}{\urlprefix }}%
\providecommand \urlprefix  [0]{URL }%
\providecommand \Eprint [0]{\href }%
\providecommand \doibase [0]{https://doi.org/}%
\providecommand \selectlanguage [0]{\@gobble}%
\providecommand \bibinfo  [0]{\@secondoftwo}%
\providecommand \bibfield  [0]{\@secondoftwo}%
\providecommand \translation [1]{[#1]}%
\providecommand \BibitemOpen [0]{}%
\providecommand \bibitemStop [0]{}%
\providecommand \bibitemNoStop [0]{.\EOS\space}%
\providecommand \EOS [0]{\spacefactor3000\relax}%
\providecommand \BibitemShut  [1]{\csname bibitem#1\endcsname}%
\let\auto@bib@innerbib\@empty
\bibitem [{\citenamefont {Bloch}\ \emph {et~al.}(2008)\citenamefont {Bloch}, \citenamefont {Dalibard},\ and\ \citenamefont {Zwerger}}]{RevModPhys.80.885}%
  \BibitemOpen
  \bibfield  {author} {\bibinfo {author} {\bibfnamefont {I.}~\bibnamefont {Bloch}}, \bibinfo {author} {\bibfnamefont {J.}~\bibnamefont {Dalibard}},\ and\ \bibinfo {author} {\bibfnamefont {W.}~\bibnamefont {Zwerger}},\ }\bibfield  {title} {\bibinfo {title} {Many-body physics with ultracold gases},\ }\href@noop {} {\bibfield  {journal} {\bibinfo  {journal} {Rev. Mod. Phys.}\ }\textbf {\bibinfo {volume} {80}},\ \bibinfo {pages} {885} (\bibinfo {year} {2008})}\BibitemShut {NoStop}%
\bibitem [{\citenamefont {Polkovnikov}\ \emph {et~al.}(2011)\citenamefont {Polkovnikov}, \citenamefont {Sengupta}, \citenamefont {Silva},\ and\ \citenamefont {Vengalattore}}]{RevModPhys.83.863}%
  \BibitemOpen
  \bibfield  {author} {\bibinfo {author} {\bibfnamefont {A.}~\bibnamefont {Polkovnikov}}, \bibinfo {author} {\bibfnamefont {K.}~\bibnamefont {Sengupta}}, \bibinfo {author} {\bibfnamefont {A.}~\bibnamefont {Silva}},\ and\ \bibinfo {author} {\bibfnamefont {M.}~\bibnamefont {Vengalattore}},\ }\bibfield  {title} {\bibinfo {title} {Colloquium: Nonequilibrium dynamics of closed interacting quantum systems},\ }\href@noop {} {\bibfield  {journal} {\bibinfo  {journal} {Rev. Mod. Phys.}\ }\textbf {\bibinfo {volume} {83}},\ \bibinfo {pages} {863} (\bibinfo {year} {2011})}\BibitemShut {NoStop}%
\bibitem [{\citenamefont {Schneider}\ \emph {et~al.}(2012)\citenamefont {Schneider}, \citenamefont {Hackermüller},\ and\ \citenamefont {Ronzheimer}}]{Schneider}%
  \BibitemOpen
  \bibfield  {author} {\bibinfo {author} {\bibfnamefont {U.}~\bibnamefont {Schneider}}, \bibinfo {author} {\bibfnamefont {L.}~\bibnamefont {Hackermüller}},\ and\ \bibinfo {author} {\bibfnamefont {J.~e.~a.}\ \bibnamefont {Ronzheimer}},\ }\bibfield  {title} {\bibinfo {title} {Fermionic transport in a homogeneous hubbard model: Out-of-equilibrium dynamics with ultracold atoms},\ }\href@noop {} {\bibfield  {journal} {\bibinfo  {journal} {Nature Phys}\ }\textbf {\bibinfo {volume} {8}},\ \bibinfo {pages} {213} (\bibinfo {year} {2012})}\BibitemShut {NoStop}%
\bibitem [{\citenamefont {Ronzheimer}\ \emph {et~al.}(2013)\citenamefont {Ronzheimer}, \citenamefont {Schreiber}, \citenamefont {Braun}, \citenamefont {Hodgman}, \citenamefont {Langer}, \citenamefont {McCulloch}, \citenamefont {Heidrich-Meisner}, \citenamefont {Bloch},\ and\ \citenamefont {Schneider}}]{Ronzheimer}%
  \BibitemOpen
  \bibfield  {author} {\bibinfo {author} {\bibfnamefont {J.~P.}\ \bibnamefont {Ronzheimer}}, \bibinfo {author} {\bibfnamefont {M.}~\bibnamefont {Schreiber}}, \bibinfo {author} {\bibfnamefont {S.}~\bibnamefont {Braun}}, \bibinfo {author} {\bibfnamefont {S.~S.}\ \bibnamefont {Hodgman}}, \bibinfo {author} {\bibfnamefont {S.}~\bibnamefont {Langer}}, \bibinfo {author} {\bibfnamefont {I.~P.}\ \bibnamefont {McCulloch}}, \bibinfo {author} {\bibfnamefont {F.}~\bibnamefont {Heidrich-Meisner}}, \bibinfo {author} {\bibfnamefont {I.}~\bibnamefont {Bloch}},\ and\ \bibinfo {author} {\bibfnamefont {U.}~\bibnamefont {Schneider}},\ }\bibfield  {title} {\bibinfo {title} {Expansion dynamics of interacting bosons in homogeneous lattices in one and two dimensions},\ }\href@noop {} {\bibfield  {journal} {\bibinfo  {journal} {Phys. Rev. Lett.}\ }\textbf {\bibinfo {volume} {110}},\ \bibinfo {pages} {205301} (\bibinfo {year} {2013})}\BibitemShut {NoStop}%
\bibitem [{\citenamefont {Collura}\ \emph {et~al.}(2013)\citenamefont {Collura}, \citenamefont {Sotiriadis},\ and\ \citenamefont {Calabrese}}]{Collura}%
  \BibitemOpen
  \bibfield  {author} {\bibinfo {author} {\bibfnamefont {M.}~\bibnamefont {Collura}}, \bibinfo {author} {\bibfnamefont {S.}~\bibnamefont {Sotiriadis}},\ and\ \bibinfo {author} {\bibfnamefont {P.}~\bibnamefont {Calabrese}},\ }\bibfield  {title} {\bibinfo {title} {Quench dynamics of a tonks–girardeau gas released from a harmonic trap},\ }\href@noop {} {\bibfield  {journal} {\bibinfo  {journal} {J. Stat. Mech.}\ }\textbf {\bibinfo {volume} {P09025}} (\bibinfo {year} {2013})}\BibitemShut {NoStop}%
\bibitem [{\citenamefont {Kinoshita}\ \emph {et~al.}()\citenamefont {Kinoshita}, \citenamefont {Wenger},\ and\ \citenamefont {Weiss}}]{science.305}%
  \BibitemOpen
  \bibfield  {author} {\bibinfo {author} {\bibfnamefont {T.}~\bibnamefont {Kinoshita}}, \bibinfo {author} {\bibfnamefont {T.}~\bibnamefont {Wenger}},\ and\ \bibinfo {author} {\bibfnamefont {D.~S.}\ \bibnamefont {Weiss}},\ }\bibfield  {title} {\bibinfo {title} {Observation of a one-dimensional tonks-girardeau gas},\ }\href@noop {} {\bibinfo  {journal} {Science}\ }\BibitemShut {NoStop}%
\bibitem [{\citenamefont {Wang}\ \emph {et~al.}(2006)\citenamefont {Wang}, \citenamefont {Fu}, \citenamefont {Liu},\ and\ \citenamefont {Wu}}]{Wang}%
  \BibitemOpen
\bibfield  {journal} {  }\bibfield  {author} {\bibinfo {author} {\bibfnamefont {B.}~\bibnamefont {Wang}}, \bibinfo {author} {\bibfnamefont {P.}~\bibnamefont {Fu}}, \bibinfo {author} {\bibfnamefont {J.}~\bibnamefont {Liu}},\ and\ \bibinfo {author} {\bibfnamefont {B.}~\bibnamefont {Wu}},\ }\bibfield  {title} {\bibinfo {title} {Self-trapping of bose-einstein condensates in optical lattices},\ }\href@noop {} {\bibfield  {journal} {\bibinfo  {journal} {Phys. Rev. A}\ }\textbf {\bibinfo {volume} {74}},\ \bibinfo {pages} {063610} (\bibinfo {year} {2006})}\BibitemShut {NoStop}%
\bibitem [{\citenamefont {Koutentakis}\ \emph {et~al.}(2017)\citenamefont {Koutentakis}, \citenamefont {Mistakidis},\ and\ \citenamefont {Schmelcher}}]{Mistakidis}%
  \BibitemOpen
  \bibfield  {author} {\bibinfo {author} {\bibfnamefont {G.~M.}\ \bibnamefont {Koutentakis}}, \bibinfo {author} {\bibfnamefont {S.~I.}\ \bibnamefont {Mistakidis}},\ and\ \bibinfo {author} {\bibfnamefont {P.}~\bibnamefont {Schmelcher}},\ }\bibfield  {title} {\bibinfo {title} {Quench-induced resonant tunneling mechanisms of bosons in an optical lattice with harmonic confinement},\ }\href@noop {} {\bibfield  {journal} {\bibinfo  {journal} {Phys. Rev. A}\ }\textbf {\bibinfo {volume} {95}},\ \bibinfo {pages} {013617} (\bibinfo {year} {2017})}\BibitemShut {NoStop}%
\bibitem [{\citenamefont {Siegl}\ \emph {et~al.}(2018)\citenamefont {Siegl}, \citenamefont {Mistakidis},\ and\ \citenamefont {Schmelcher}}]{PhysRevA.97.053626}%
  \BibitemOpen
  \bibfield  {author} {\bibinfo {author} {\bibfnamefont {P.}~\bibnamefont {Siegl}}, \bibinfo {author} {\bibfnamefont {S.~I.}\ \bibnamefont {Mistakidis}},\ and\ \bibinfo {author} {\bibfnamefont {P.}~\bibnamefont {Schmelcher}},\ }\bibfield  {title} {\bibinfo {title} {Many-body expansion dynamics of a bose-fermi mixture confined in an optical lattice},\ }\href@noop {} {\bibfield  {journal} {\bibinfo  {journal} {Phys. Rev. A}\ }\textbf {\bibinfo {volume} {97}},\ \bibinfo {pages} {053626} (\bibinfo {year} {2018})}\BibitemShut {NoStop}%
\bibitem [{\citenamefont {Vidmar}\ \emph {et~al.}(2015)\citenamefont {Vidmar}, \citenamefont {Ronzheimer}, \citenamefont {Schreiber}, \citenamefont {Braun}, \citenamefont {Langer}, \citenamefont {Bloch},\ and\ \citenamefont {Schneider}}]{Langer}%
  \BibitemOpen
  \bibfield  {author} {\bibinfo {author} {\bibfnamefont {L.}~\bibnamefont {Vidmar}}, \bibinfo {author} {\bibfnamefont {J.}~\bibnamefont {Ronzheimer}}, \bibinfo {author} {\bibfnamefont {M.}~\bibnamefont {Schreiber}}, \bibinfo {author} {\bibfnamefont {S.}~\bibnamefont {Braun}}, \bibinfo {author} {\bibfnamefont {S.~H.~S.}\ \bibnamefont {Langer}}, \bibinfo {author} {\bibfnamefont {F.~H.-M.~I.}\ \bibnamefont {Bloch}},\ and\ \bibinfo {author} {\bibfnamefont {U.}~\bibnamefont {Schneider}},\ }\bibfield  {title} {\bibinfo {title} {Dynamical quasicondensation of hard-core bosons at finite momenta},\ }\href@noop {} {\bibfield  {journal} {\bibinfo  {journal} {Phys. Rev. Lett.}\ }\textbf {\bibinfo {volume} {115}},\ \bibinfo {pages} {175301} (\bibinfo {year} {2015})}\BibitemShut {NoStop}%
\bibitem [{\citenamefont {Reinhard}\ \emph {et~al.}(2013)\citenamefont {Reinhard}, \citenamefont {Riou}, \citenamefont {Zundel}, \citenamefont {Weiss}, \citenamefont {Li}, \citenamefont {Rey},\ and\ \citenamefont {Hipolito}}]{Weiss}%
  \BibitemOpen
  \bibfield  {author} {\bibinfo {author} {\bibfnamefont {A.}~\bibnamefont {Reinhard}}, \bibinfo {author} {\bibfnamefont {J.-F.}\ \bibnamefont {Riou}}, \bibinfo {author} {\bibfnamefont {L.~A.}\ \bibnamefont {Zundel}}, \bibinfo {author} {\bibfnamefont {D.~S.}\ \bibnamefont {Weiss}}, \bibinfo {author} {\bibfnamefont {S.}~\bibnamefont {Li}}, \bibinfo {author} {\bibfnamefont {A.~M.}\ \bibnamefont {Rey}},\ and\ \bibinfo {author} {\bibfnamefont {R.}~\bibnamefont {Hipolito}},\ }\bibfield  {title} {\bibinfo {title} {Self-trapping in an array of coupled 1d bose gases},\ }\href@noop {} {\bibfield  {journal} {\bibinfo  {journal} {Phys. Rev. Lett.}\ }\textbf {\bibinfo {volume} {110}},\ \bibinfo {pages} {033001} (\bibinfo {year} {2013})}\BibitemShut {NoStop}%
\bibitem [{\citenamefont {Xia}\ \emph {et~al.}(2015)\citenamefont {Xia}, \citenamefont {Zundel}, \citenamefont {Carrasquilla}, \citenamefont {Reinhard}, \citenamefont {Wilson}, \citenamefont {Rigol},\ and\ \citenamefont {Weiss}}]{Weiss1}%
  \BibitemOpen
  \bibfield  {author} {\bibinfo {author} {\bibfnamefont {L.}~\bibnamefont {Xia}}, \bibinfo {author} {\bibfnamefont {L.~A.}\ \bibnamefont {Zundel}}, \bibinfo {author} {\bibfnamefont {J.}~\bibnamefont {Carrasquilla}}, \bibinfo {author} {\bibfnamefont {A.}~\bibnamefont {Reinhard}}, \bibinfo {author} {\bibfnamefont {J.~M.}\ \bibnamefont {Wilson}}, \bibinfo {author} {\bibfnamefont {M.}~\bibnamefont {Rigol}},\ and\ \bibinfo {author} {\bibfnamefont {D.~S.}\ \bibnamefont {Weiss}},\ }\bibfield  {title} {\bibinfo {title} {Quantum distillation and confinement of vacancies in a doublon sea},\ }\href@noop {} {\bibfield  {journal} {\bibinfo  {journal} {Nature Phys}\ }\textbf {\bibinfo {volume} {11}},\ \bibinfo {pages} {316} (\bibinfo {year} {2015})}\BibitemShut {NoStop}%
\bibitem [{\citenamefont {Xu}\ and\ \citenamefont {Rigol}(2017)}]{Weixu}%
  \BibitemOpen
  \bibfield  {author} {\bibinfo {author} {\bibfnamefont {W.}~\bibnamefont {Xu}}\ and\ \bibinfo {author} {\bibfnamefont {M.}~\bibnamefont {Rigol}},\ }\bibfield  {title} {\bibinfo {title} {Expansion of one-dimensional lattice hard-core bosons at finite temperature},\ }\href@noop {} {\bibfield  {journal} {\bibinfo  {journal} {Phys. Rev. A}\ }\textbf {\bibinfo {volume} {95}},\ \bibinfo {pages} {033617} (\bibinfo {year} {2017})}\BibitemShut {NoStop}%
\bibitem [{\citenamefont {Trujillo-Martinez}\ \emph {et~al.}(2021)\citenamefont {Trujillo-Martinez}, \citenamefont {Posazhennikova},\ and\ \citenamefont {Kroha}}]{Anna}%
  \BibitemOpen
  \bibfield  {author} {\bibinfo {author} {\bibfnamefont {M.}~\bibnamefont {Trujillo-Martinez}}, \bibinfo {author} {\bibfnamefont {A.}~\bibnamefont {Posazhennikova}},\ and\ \bibinfo {author} {\bibfnamefont {J.}~\bibnamefont {Kroha}},\ }\bibfield  {title} {\bibinfo {title} {Expansion dynamics in two-dimensional bose-hubbard lattices: Bose-einstein condensate and thermal cloud},\ }\href@noop {} {\bibfield  {journal} {\bibinfo  {journal} {Phys. Rev. A}\ }\textbf {\bibinfo {volume} {103}},\ \bibinfo {pages} {033311} (\bibinfo {year} {2021})}\BibitemShut {NoStop}%
\bibitem [{\citenamefont {Boschi}\ \emph {et~al.}(2014)\citenamefont {Boschi}, \citenamefont {Ercolessi}, \citenamefont {Ferrari}, \citenamefont {Naldesi}, \citenamefont {Ortolani},\ and\ \citenamefont {Taddia}}]{Loris}%
  \BibitemOpen
  \bibfield  {author} {\bibinfo {author} {\bibfnamefont {C.~D.~E.}\ \bibnamefont {Boschi}}, \bibinfo {author} {\bibfnamefont {E.}~\bibnamefont {Ercolessi}}, \bibinfo {author} {\bibfnamefont {L.}~\bibnamefont {Ferrari}}, \bibinfo {author} {\bibfnamefont {P.}~\bibnamefont {Naldesi}}, \bibinfo {author} {\bibfnamefont {F.}~\bibnamefont {Ortolani}},\ and\ \bibinfo {author} {\bibfnamefont {L.}~\bibnamefont {Taddia}},\ }\bibfield  {title} {\bibinfo {title} {Bound states and expansion dynamics of interacting bosons on a one-dimensional lattice},\ }\href@noop {} {\bibfield  {journal} {\bibinfo  {journal} {Phys. Rev. A}\ }\textbf {\bibinfo {volume} {90}},\ \bibinfo {pages} {043606} (\bibinfo {year} {2014})}\BibitemShut {NoStop}%
\bibitem [{\citenamefont {Mark~Jreissaty}\ and\ \citenamefont {Rigol}(2011)}]{Wolf}%
  \BibitemOpen
  \bibfield  {author} {\bibinfo {author} {\bibfnamefont {F.~A.~W.}\ \bibnamefont {Mark~Jreissaty}, \bibfnamefont {Juan~Carrasquilla}}\ and\ \bibinfo {author} {\bibfnamefont {M.}~\bibnamefont {Rigol}},\ }\bibfield  {title} {\bibinfo {title} {Expansion of bose-hubbard mott insulators in optical lattices},\ }\href@noop {} {\bibfield  {journal} {\bibinfo  {journal} {Phys. Rev. A}\ }\textbf {\bibinfo {volume} {84}},\ \bibinfo {pages} {043610} (\bibinfo {year} {2011})}\BibitemShut {NoStop}%
\bibitem [{\citenamefont {Jreissaty}\ \emph {et~al.}(2013)\citenamefont {Jreissaty}, \citenamefont {Carrasquilla},\ and\ \citenamefont {Rigol}}]{Juan}%
  \BibitemOpen
  \bibfield  {author} {\bibinfo {author} {\bibfnamefont {A.}~\bibnamefont {Jreissaty}}, \bibinfo {author} {\bibfnamefont {J.}~\bibnamefont {Carrasquilla}},\ and\ \bibinfo {author} {\bibfnamefont {M.}~\bibnamefont {Rigol}},\ }\bibfield  {title} {\bibinfo {title} {Self-trapping in the two-dimensional bose-hubbard model},\ }\href@noop {} {\bibfield  {journal} {\bibinfo  {journal} {Phys. Rev. A}\ }\textbf {\bibinfo {volume} {88}},\ \bibinfo {pages} {031606(R)} (\bibinfo {year} {2013})}\BibitemShut {NoStop}%
\bibitem [{\citenamefont {L.~Vidmar}\ \emph {et~al.}(2013)\citenamefont {L.~Vidmar}, \citenamefont {McCulloch}, \citenamefont {Schneider}, \citenamefont {Schollwöck},\ and\ \citenamefont {Heidrich-Meisner}}]{Langer1}%
  \BibitemOpen
  \bibfield  {author} {\bibinfo {author} {\bibfnamefont {S.~L.}\ \bibnamefont {L.~Vidmar}}, \bibinfo {author} {\bibfnamefont {I.~P.}\ \bibnamefont {McCulloch}}, \bibinfo {author} {\bibfnamefont {U.}~\bibnamefont {Schneider}}, \bibinfo {author} {\bibfnamefont {U.}~\bibnamefont {Schollwöck}},\ and\ \bibinfo {author} {\bibfnamefont {F.}~\bibnamefont {Heidrich-Meisner}},\ }\bibfield  {title} {\bibinfo {title} {Sudden expansion of mott insulators in one dimension},\ }\href@noop {} {\bibfield  {journal} {\bibinfo  {journal} {Phys. Rev. B}\ }\textbf {\bibinfo {volume} {88}},\ \bibinfo {pages} {235117} (\bibinfo {year} {2013})}\BibitemShut {NoStop}%
\bibitem [{\citenamefont {Hauschild}\ \emph {et~al.}(2015)\citenamefont {Hauschild}, \citenamefont {Pollmann},\ and\ \citenamefont {Heidrich-Meisner}}]{Frank}%
  \BibitemOpen
  \bibfield  {author} {\bibinfo {author} {\bibfnamefont {J.}~\bibnamefont {Hauschild}}, \bibinfo {author} {\bibfnamefont {F.}~\bibnamefont {Pollmann}},\ and\ \bibinfo {author} {\bibfnamefont {F.}~\bibnamefont {Heidrich-Meisner}},\ }\bibfield  {title} {\bibinfo {title} {Sudden expansion and domain-wall melting of strongly interacting bosons in two-dimensional optical lattices and on multileg ladders},\ }\href@noop {} {\bibfield  {journal} {\bibinfo  {journal} {Phys. Rev. A}\ }\textbf {\bibinfo {volume} {92}},\ \bibinfo {pages} {053629} (\bibinfo {year} {2015})}\BibitemShut {NoStop}%
\bibitem [{\citenamefont {Anker}\ \emph {et~al.}(2005)\citenamefont {Anker}, \citenamefont {Albiez}, \citenamefont {Gati}, \citenamefont {Hunsmann}, \citenamefont {Eiermann}, \citenamefont {Trombettoni},\ and\ \citenamefont {Oberthaler}}]{Anker}%
  \BibitemOpen
  \bibfield  {author} {\bibinfo {author} {\bibfnamefont {T.}~\bibnamefont {Anker}}, \bibinfo {author} {\bibfnamefont {M.}~\bibnamefont {Albiez}}, \bibinfo {author} {\bibfnamefont {R.}~\bibnamefont {Gati}}, \bibinfo {author} {\bibfnamefont {S.}~\bibnamefont {Hunsmann}}, \bibinfo {author} {\bibfnamefont {B.}~\bibnamefont {Eiermann}}, \bibinfo {author} {\bibfnamefont {A.}~\bibnamefont {Trombettoni}},\ and\ \bibinfo {author} {\bibfnamefont {M.~K.}\ \bibnamefont {Oberthaler}},\ }\bibfield  {title} {\bibinfo {title} {Nonlinear self-trapping of matter waves in periodic potentials},\ }\href@noop {} {\bibfield  {journal} {\bibinfo  {journal} {Phys. Rev. Lett.}\ }\textbf {\bibinfo {volume} {94}},\ \bibinfo {pages} {020403} (\bibinfo {year} {2005})}\BibitemShut {NoStop}%
\bibitem [{\citenamefont {Albiez}\ \emph {et~al.}(2005)\citenamefont {Albiez}, \citenamefont {Gati}, \citenamefont {F\"olling}, \citenamefont {Hunsmann}, \citenamefont {Cristiani},\ and\ \citenamefont {Oberthaler}}]{Albiez05}%
  \BibitemOpen
  \bibfield  {author} {\bibinfo {author} {\bibfnamefont {M.}~\bibnamefont {Albiez}}, \bibinfo {author} {\bibfnamefont {R.}~\bibnamefont {Gati}}, \bibinfo {author} {\bibfnamefont {J.}~\bibnamefont {F\"olling}}, \bibinfo {author} {\bibfnamefont {S.}~\bibnamefont {Hunsmann}}, \bibinfo {author} {\bibfnamefont {M.}~\bibnamefont {Cristiani}},\ and\ \bibinfo {author} {\bibfnamefont {M.~K.}\ \bibnamefont {Oberthaler}},\ }\bibfield  {title} {\bibinfo {title} {Direct observation of tunneling and nonlinear self-trapping in a single bosonic josephson junction},\ }\href {https://doi.org/10.1103/PhysRevLett.95.010402} {\bibfield  {journal} {\bibinfo  {journal} {Phys. Rev. Lett.}\ }\textbf {\bibinfo {volume} {95}},\ \bibinfo {pages} {010402} (\bibinfo {year} {2005})}\BibitemShut {NoStop}%
\bibitem [{\citenamefont {Smerzi}\ \emph {et~al.}(1997)\citenamefont {Smerzi}, \citenamefont {Fantoni}, \citenamefont {Giovanazzi},\ and\ \citenamefont {Shenoy}}]{Smerzi97}%
  \BibitemOpen
  \bibfield  {author} {\bibinfo {author} {\bibfnamefont {A.}~\bibnamefont {Smerzi}}, \bibinfo {author} {\bibfnamefont {S.}~\bibnamefont {Fantoni}}, \bibinfo {author} {\bibfnamefont {S.}~\bibnamefont {Giovanazzi}},\ and\ \bibinfo {author} {\bibfnamefont {S.~R.}\ \bibnamefont {Shenoy}},\ }\bibfield  {title} {\bibinfo {title} {Quantum coherent atomic tunneling between two trapped bose-einstein condensates},\ }\href {https://doi.org/10.1103/PhysRevLett.79.4950} {\bibfield  {journal} {\bibinfo  {journal} {Phys. Rev. Lett.}\ }\textbf {\bibinfo {volume} {79}},\ \bibinfo {pages} {4950} (\bibinfo {year} {1997})}\BibitemShut {NoStop}%
\bibitem [{\citenamefont {Raghavan}\ \emph {et~al.}(1999)\citenamefont {Raghavan}, \citenamefont {Smerzi}, \citenamefont {Fantoni},\ and\ \citenamefont {Shenoy}}]{Raghavan99}%
  \BibitemOpen
  \bibfield  {author} {\bibinfo {author} {\bibfnamefont {S.}~\bibnamefont {Raghavan}}, \bibinfo {author} {\bibfnamefont {A.}~\bibnamefont {Smerzi}}, \bibinfo {author} {\bibfnamefont {S.}~\bibnamefont {Fantoni}},\ and\ \bibinfo {author} {\bibfnamefont {S.~R.}\ \bibnamefont {Shenoy}},\ }\bibfield  {title} {\bibinfo {title} {Coherent oscillations between two weakly coupled bose-einstein condensates: Josephson effects, $\ensuremath{\pi}$ oscillations, and macroscopic quantum self-trapping},\ }\href {https://doi.org/10.1103/PhysRevA.59.620} {\bibfield  {journal} {\bibinfo  {journal} {Phys. Rev. A}\ }\textbf {\bibinfo {volume} {59}},\ \bibinfo {pages} {620} (\bibinfo {year} {1999})}\BibitemShut {NoStop}%
\bibitem [{\citenamefont {Trombettoni}\ and\ \citenamefont {Smerzi}(2001{\natexlab{a}})}]{Trombettoni01}%
  \BibitemOpen
  \bibfield  {author} {\bibinfo {author} {\bibfnamefont {A.}~\bibnamefont {Trombettoni}}\ and\ \bibinfo {author} {\bibfnamefont {A.}~\bibnamefont {Smerzi}},\ }\bibfield  {title} {\bibinfo {title} {Discrete solitons and breathers with dilute bose-einstein condensates},\ }\href {https://doi.org/10.1103/PhysRevLett.86.2353} {\bibfield  {journal} {\bibinfo  {journal} {Phys. Rev. Lett.}\ }\textbf {\bibinfo {volume} {86}},\ \bibinfo {pages} {2353} (\bibinfo {year} {2001}{\natexlab{a}})}\BibitemShut {NoStop}%
\bibitem [{\citenamefont {Trombettoni}\ and\ \citenamefont {Smerzi}(2001{\natexlab{b}})}]{Trombettoni_2001}%
  \BibitemOpen
  \bibfield  {author} {\bibinfo {author} {\bibfnamefont {A.}~\bibnamefont {Trombettoni}}\ and\ \bibinfo {author} {\bibfnamefont {A.}~\bibnamefont {Smerzi}},\ }\bibfield  {title} {\bibinfo {title} {Variational dynamics of bose-einstein condensates in deep optical lattices},\ }\href {https://doi.org/10.1088/0953-4075/34/23/315} {\bibfield  {journal} {\bibinfo  {journal} {Journal of Physics B: Atomic, Molecular and Optical Physics}\ }\textbf {\bibinfo {volume} {34}},\ \bibinfo {pages} {4711} (\bibinfo {year} {2001}{\natexlab{b}})}\BibitemShut {NoStop}%
\bibitem [{\citenamefont {Creffield}(2007)}]{Creffield}%
  \BibitemOpen
  \bibfield  {author} {\bibinfo {author} {\bibfnamefont {C.~E.}\ \bibnamefont {Creffield}},\ }\bibfield  {title} {\bibinfo {title} {Coherent control of self-trapping of cold bosonic atoms},\ }\href@noop {} {\bibfield  {journal} {\bibinfo  {journal} {Phys. Rev. A}\ }\textbf {\bibinfo {volume} {75}},\ \bibinfo {pages} {031607(R)} (\bibinfo {year} {2007})}\BibitemShut {NoStop}%
\bibitem [{\citenamefont {Rosenkranz}\ \emph {et~al.}(2007)\citenamefont {Rosenkranz}, \citenamefont {Jaksch}, \citenamefont {Lim},\ and\ \citenamefont {Ba}}]{Bao}%
  \BibitemOpen
  \bibfield  {author} {\bibinfo {author} {\bibfnamefont {M.}~\bibnamefont {Rosenkranz}}, \bibinfo {author} {\bibfnamefont {D.}~\bibnamefont {Jaksch}}, \bibinfo {author} {\bibfnamefont {F.~Y.}\ \bibnamefont {Lim}},\ and\ \bibinfo {author} {\bibfnamefont {W.}~\bibnamefont {Ba}},\ }\bibfield  {title} {\bibinfo {title} {Self-trapping of bose-einstein condensates expanding into shallow optical lattices},\ }\href@noop {} {\bibfield  {journal} {\bibinfo  {journal} {Phys. Rev. A}\ }\textbf {\bibinfo {volume} {77}},\ \bibinfo {pages} {063607} (\bibinfo {year} {2007})}\BibitemShut {NoStop}%
\bibitem [{\citenamefont {Xiong}\ \emph {et~al.}(2009)\citenamefont {Xiong}, \citenamefont {Gong}, \citenamefont {Pu}, \citenamefont {Bao},\ and\ \citenamefont {Li}}]{Bo}%
  \BibitemOpen
  \bibfield  {author} {\bibinfo {author} {\bibfnamefont {B.}~\bibnamefont {Xiong}}, \bibinfo {author} {\bibfnamefont {J.}~\bibnamefont {Gong}}, \bibinfo {author} {\bibfnamefont {H.}~\bibnamefont {Pu}}, \bibinfo {author} {\bibfnamefont {W.}~\bibnamefont {Bao}},\ and\ \bibinfo {author} {\bibfnamefont {B.}~\bibnamefont {Li}},\ }\bibfield  {title} {\bibinfo {title} {Symmetry breaking and self-trapping of a dipolar bose-einstein condensate in a double-well potential},\ }\href@noop {} {\bibfield  {journal} {\bibinfo  {journal} {Phys. Rev. A}\ }\textbf {\bibinfo {volume} {79}},\ \bibinfo {pages} {013626} (\bibinfo {year} {2009})}\BibitemShut {NoStop}%
\bibitem [{\citenamefont {Roy}\ \emph {et~al.}(2022)\citenamefont {Roy}, \citenamefont {Chakrabarti},\ and\ \citenamefont {Trombettoni}}]{rhombik_epjd}%
  \BibitemOpen
  \bibfield  {author} {\bibinfo {author} {\bibfnamefont {R.}~\bibnamefont {Roy}}, \bibinfo {author} {\bibfnamefont {B.}~\bibnamefont {Chakrabarti}},\ and\ \bibinfo {author} {\bibfnamefont {A.}~\bibnamefont {Trombettoni}},\ }\bibfield  {title} {\bibinfo {title} {Quantum dynamics of few dipolar bosons in a double-well potential.},\ }\href@noop {} {\bibfield  {journal} {\bibinfo  {journal} {European Physical Journal D}\ }\textbf {\bibinfo {volume} {76}},\ \bibinfo {pages} {215303} (\bibinfo {year} {2022})}\BibitemShut {NoStop}%
\bibitem [{\citenamefont {Morsch}\ and\ \citenamefont {Oberthaler}(2006)}]{Oberthaler06}%
  \BibitemOpen
  \bibfield  {author} {\bibinfo {author} {\bibfnamefont {O.}~\bibnamefont {Morsch}}\ and\ \bibinfo {author} {\bibfnamefont {M.}~\bibnamefont {Oberthaler}},\ }\bibfield  {title} {\bibinfo {title} {Dynamics of bose-einstein condensates in optical lattices},\ }\href {https://doi.org/10.1103/RevModPhys.78.179} {\bibfield  {journal} {\bibinfo  {journal} {Rev. Mod. Phys.}\ }\textbf {\bibinfo {volume} {78}},\ \bibinfo {pages} {179} (\bibinfo {year} {2006})}\BibitemShut {NoStop}%
\bibitem [{\citenamefont {Duan}(2005)}]{Duan}%
  \BibitemOpen
  \bibfield  {author} {\bibinfo {author} {\bibfnamefont {L.~M.}\ \bibnamefont {Duan}},\ }\bibfield  {title} {\bibinfo {title} {Effective hamiltonian for fermions in an optical lattice across a feshbach resonance},\ }\href@noop {} {\bibfield  {journal} {\bibinfo  {journal} {Phys. Rev. Lett.}\ }\textbf {\bibinfo {volume} {95}},\ \bibinfo {pages} {243202} (\bibinfo {year} {2005})}\BibitemShut {NoStop}%
\bibitem [{\citenamefont {Li}\ \emph {et~al.}(2012)\citenamefont {Li}, \citenamefont {Hamadeh},\ and\ \citenamefont {Lesanovsky}}]{Lama}%
  \BibitemOpen
  \bibfield  {author} {\bibinfo {author} {\bibfnamefont {W.}~\bibnamefont {Li}}, \bibinfo {author} {\bibfnamefont {L.}~\bibnamefont {Hamadeh}},\ and\ \bibinfo {author} {\bibfnamefont {I.}~\bibnamefont {Lesanovsky}},\ }\bibfield  {title} {\bibinfo {title} {Probing the interaction between rydberg-dressed atoms through interference},\ }\href@noop {} {\bibfield  {journal} {\bibinfo  {journal} {Phys. Rev. A}\ }\textbf {\bibinfo {volume} {85}},\ \bibinfo {pages} {053615} (\bibinfo {year} {2012})}\BibitemShut {NoStop}%
\bibitem [{\citenamefont {Defenu}\ \emph {et~al.}(2023)\citenamefont {Defenu}, \citenamefont {Donner}, \citenamefont {Macr\`{\i}}, \citenamefont {Pagano}, \citenamefont {Ruffo},\ and\ \citenamefont {Trombettoni}}]{Defenu23}%
  \BibitemOpen
  \bibfield  {author} {\bibinfo {author} {\bibfnamefont {N.}~\bibnamefont {Defenu}}, \bibinfo {author} {\bibfnamefont {T.}~\bibnamefont {Donner}}, \bibinfo {author} {\bibfnamefont {T.}~\bibnamefont {Macr\`{\i}}}, \bibinfo {author} {\bibfnamefont {G.}~\bibnamefont {Pagano}}, \bibinfo {author} {\bibfnamefont {S.}~\bibnamefont {Ruffo}},\ and\ \bibinfo {author} {\bibfnamefont {A.}~\bibnamefont {Trombettoni}},\ }\bibfield  {title} {\bibinfo {title} {Long-range interacting quantum systems},\ }\href {https://doi.org/10.1103/RevModPhys.95.035002} {\bibfield  {journal} {\bibinfo  {journal} {Rev. Mod. Phys.}\ }\textbf {\bibinfo {volume} {95}},\ \bibinfo {pages} {035002} (\bibinfo {year} {2023})}\BibitemShut {NoStop}%
\bibitem [{\citenamefont {Hazzard}\ \emph {et~al.}(2014)\citenamefont {Hazzard}, \citenamefont {Gadway},\ and\ \citenamefont {Michael Foss-Feig}}]{Kaden}%
  \BibitemOpen
  \bibfield  {author} {\bibinfo {author} {\bibfnamefont {K.~R.}\ \bibnamefont {Hazzard}}, \bibinfo {author} {\bibfnamefont {B.}~\bibnamefont {Gadway}},\ and\ \bibinfo {author} {\bibfnamefont {e.~a.}\ \bibnamefont {Michael Foss-Feig}},\ }\bibfield  {title} {\bibinfo {title} {Many-body dynamics of dipolar molecules in an optical lattice},\ }\href@noop {} {\bibfield  {journal} {\bibinfo  {journal} {Phys. Rev. Lett.}\ }\textbf {\bibinfo {volume} {113}},\ \bibinfo {pages} {195302} (\bibinfo {year} {2014})}\BibitemShut {NoStop}%
\bibitem [{\citenamefont {T.}\ \emph {et~al.}(2007)\citenamefont {T.}, \citenamefont {Koch},\ and\ \citenamefont {et~al.}}]{Koch}%
  \BibitemOpen
  \bibfield  {author} {\bibinfo {author} {\bibfnamefont {L.}~\bibnamefont {T.}}, \bibinfo {author} {\bibfnamefont {T.}~\bibnamefont {Koch}},\ and\ \bibinfo {author} {\bibfnamefont {F.~B.}\ \bibnamefont {et~al.}},\ }\bibfield  {title} {\bibinfo {title} {Strong dipolar effects in a quantum ferrofluid.},\ }\href@noop {} {\bibfield  {journal} {\bibinfo  {journal} {Nature}\ }\textbf {\bibinfo {volume} {448}},\ \bibinfo {pages} {672} (\bibinfo {year} {2007})}\BibitemShut {NoStop}%
\bibitem [{\citenamefont {Lahaye}\ \emph {et~al.}(2008)\citenamefont {Lahaye}, \citenamefont {Metz}, \citenamefont {ans T.~Koch}, \citenamefont {Meister}, \citenamefont {Griesmaier}, \citenamefont {Pfau}, \citenamefont {H.~Saito},\ and\ \citenamefont {Ueda}}]{Koch1}%
  \BibitemOpen
  \bibfield  {author} {\bibinfo {author} {\bibfnamefont {T.}~\bibnamefont {Lahaye}}, \bibinfo {author} {\bibfnamefont {J.}~\bibnamefont {Metz}}, \bibinfo {author} {\bibfnamefont {B.~F.}\ \bibnamefont {ans T.~Koch}}, \bibinfo {author} {\bibfnamefont {M.}~\bibnamefont {Meister}}, \bibinfo {author} {\bibfnamefont {A.}~\bibnamefont {Griesmaier}}, \bibinfo {author} {\bibfnamefont {T.}~\bibnamefont {Pfau}}, \bibinfo {author} {\bibfnamefont {Y.~K.}\ \bibnamefont {H.~Saito}},\ and\ \bibinfo {author} {\bibfnamefont {M.}~\bibnamefont {Ueda}},\ }\bibfield  {title} {\bibinfo {title} {d-wave collapse and explosion of a dipolar bose-einstein condensate},\ }\href@noop {} {\bibfield  {journal} {\bibinfo  {journal} {Phys. Rev. Lett.}\ }\textbf {\bibinfo {volume} {101}},\ \bibinfo {pages} {080401} (\bibinfo {year} {2008})}\BibitemShut {NoStop}%
\bibitem [{\citenamefont {Koch}\ \emph {et~al.}(2008)\citenamefont {Koch}, \citenamefont {Lahaye},\ and\ \citenamefont {et~al.}}]{Koch2}%
  \BibitemOpen
  \bibfield  {author} {\bibinfo {author} {\bibfnamefont {T.}~\bibnamefont {Koch}}, \bibinfo {author} {\bibfnamefont {T.}~\bibnamefont {Lahaye}},\ and\ \bibinfo {author} {\bibfnamefont {M.~J.}\ \bibnamefont {et~al.}},\ }\bibfield  {title} {\bibinfo {title} {Stabilization of a purely dipolar quantum gas against collapse.},\ }\href@noop {} {\bibfield  {journal} {\bibinfo  {journal} {Nature Phys}\ }\textbf {\bibinfo {volume} {4}},\ \bibinfo {pages} {218} (\bibinfo {year} {2008})}\BibitemShut {NoStop}%
\bibitem [{\citenamefont {Griesmaier}\ \emph {et~al.}(2005)\citenamefont {Griesmaier}, \citenamefont {Werner}, \citenamefont {Hensler}, \citenamefont {Stuhler},\ and\ \citenamefont {Pfau}}]{cro}%
  \BibitemOpen
  \bibfield  {author} {\bibinfo {author} {\bibfnamefont {A.}~\bibnamefont {Griesmaier}}, \bibinfo {author} {\bibfnamefont {J.}~\bibnamefont {Werner}}, \bibinfo {author} {\bibfnamefont {S.}~\bibnamefont {Hensler}}, \bibinfo {author} {\bibfnamefont {J.}~\bibnamefont {Stuhler}},\ and\ \bibinfo {author} {\bibfnamefont {T.}~\bibnamefont {Pfau}},\ }\bibfield  {title} {\bibinfo {title} {Bose-einstein condensation of chromium},\ }\href@noop {} {\bibfield  {journal} {\bibinfo  {journal} {Phys. Rev. Lett.}\ }\textbf {\bibinfo {volume} {94}},\ \bibinfo {pages} {160401} (\bibinfo {year} {2005})}\BibitemShut {NoStop}%
\bibitem [{\citenamefont {Lu}\ \emph {et~al.}(2011)\citenamefont {Lu}, \citenamefont {Burdick}, \citenamefont {Youn},\ and\ \citenamefont {Lev}}]{dys}%
  \BibitemOpen
  \bibfield  {author} {\bibinfo {author} {\bibfnamefont {M.}~\bibnamefont {Lu}}, \bibinfo {author} {\bibfnamefont {N.~Q.}\ \bibnamefont {Burdick}}, \bibinfo {author} {\bibfnamefont {S.~H.}\ \bibnamefont {Youn}},\ and\ \bibinfo {author} {\bibfnamefont {B.~L.}\ \bibnamefont {Lev}},\ }\bibfield  {title} {\bibinfo {title} {Strongly dipolar bose-einstein condensate of dysprosium},\ }\href@noop {} {\bibfield  {journal} {\bibinfo  {journal} {Phys. Rev. Lett.}\ }\textbf {\bibinfo {volume} {107}},\ \bibinfo {pages} {190401} (\bibinfo {year} {2011})}\BibitemShut {NoStop}%
\bibitem [{\citenamefont {Aikawa}\ \emph {et~al.}(2012)\citenamefont {Aikawa}, \citenamefont {Frisch}, \citenamefont {Mark}, \citenamefont {Baier}, \citenamefont {Rietzler}, \citenamefont {Grimm},\ and\ \citenamefont {Ferlaino}}]{erb}%
  \BibitemOpen
  \bibfield  {author} {\bibinfo {author} {\bibfnamefont {K.}~\bibnamefont {Aikawa}}, \bibinfo {author} {\bibfnamefont {A.}~\bibnamefont {Frisch}}, \bibinfo {author} {\bibfnamefont {M.}~\bibnamefont {Mark}}, \bibinfo {author} {\bibfnamefont {S.}~\bibnamefont {Baier}}, \bibinfo {author} {\bibfnamefont {A.}~\bibnamefont {Rietzler}}, \bibinfo {author} {\bibfnamefont {R.}~\bibnamefont {Grimm}},\ and\ \bibinfo {author} {\bibfnamefont {F.}~\bibnamefont {Ferlaino}},\ }\bibfield  {title} {\bibinfo {title} {Bose-einstein condensation of erbium},\ }\href@noop {} {\bibfield  {journal} {\bibinfo  {journal} {Phys. Rev. Lett.}\ }\textbf {\bibinfo {volume} {108}},\ \bibinfo {pages} {210401} (\bibinfo {year} {2012})}\BibitemShut {NoStop}%
\bibitem [{\citenamefont {Böttcher}\ \emph {et~al.}(2021)\citenamefont {Böttcher}, \citenamefont {Schmidt}, \citenamefont {Hertkorn}, \citenamefont {Ng}, \citenamefont {Graham1}, \citenamefont {Guo}, \citenamefont {Langen},\ and\ \citenamefont {Pfau}}]{ref1}%
  \BibitemOpen
  \bibfield  {author} {\bibinfo {author} {\bibfnamefont {F.}~\bibnamefont {Böttcher}}, \bibinfo {author} {\bibfnamefont {J.-N.}\ \bibnamefont {Schmidt}}, \bibinfo {author} {\bibfnamefont {J.}~\bibnamefont {Hertkorn}}, \bibinfo {author} {\bibfnamefont {K.~S.~H.}\ \bibnamefont {Ng}}, \bibinfo {author} {\bibfnamefont {S.~D.}\ \bibnamefont {Graham1}}, \bibinfo {author} {\bibfnamefont {M.}~\bibnamefont {Guo}}, \bibinfo {author} {\bibfnamefont {T.}~\bibnamefont {Langen}},\ and\ \bibinfo {author} {\bibfnamefont {T.}~\bibnamefont {Pfau}},\ }\bibfield  {title} {\bibinfo {title} {New states of matter with fine-tuned interactions: quantum droplets and dipolar supersolids},\ }\href@noop {} {\bibfield  {journal} {\bibinfo  {journal} {Rep. Prog. Phys.}\ }\textbf {\bibinfo {volume} {84}},\ \bibinfo {pages} {012403} (\bibinfo {year} {2021})}\BibitemShut {NoStop}%
\bibitem [{\citenamefont {Chomaz}\ \emph {et~al.}(2023)\citenamefont {Chomaz}, \citenamefont {Ferrier-Barbut}, \citenamefont {Ferlaino}, \citenamefont {Laburthe-Tolra}, \citenamefont {Lev},\ and\ \citenamefont {Pfau}}]{ref2}%
  \BibitemOpen
  \bibfield  {author} {\bibinfo {author} {\bibfnamefont {L.}~\bibnamefont {Chomaz}}, \bibinfo {author} {\bibfnamefont {I.}~\bibnamefont {Ferrier-Barbut}}, \bibinfo {author} {\bibfnamefont {F.}~\bibnamefont {Ferlaino}}, \bibinfo {author} {\bibfnamefont {B.}~\bibnamefont {Laburthe-Tolra}}, \bibinfo {author} {\bibfnamefont {B.~L.}\ \bibnamefont {Lev}},\ and\ \bibinfo {author} {\bibfnamefont {T.}~\bibnamefont {Pfau}},\ }\bibfield  {title} {\bibinfo {title} {Dipolar physics: a review of experiments with magnetic quantum gases},\ }\href@noop {} {\bibfield  {journal} {\bibinfo  {journal} {Rep. Prog. Phys.}\ }\textbf {\bibinfo {volume} {86}},\ \bibinfo {pages} {026401} (\bibinfo {year} {2023})}\BibitemShut {NoStop}%
\bibitem [{\citenamefont {Su}\ \emph {et~al.}(2023)\citenamefont {Su}, \citenamefont {Douglas},\ and\ \citenamefont {Szurek}}]{Lin:2023}%
  \BibitemOpen
  \bibfield  {author} {\bibinfo {author} {\bibfnamefont {L.}~\bibnamefont {Su}}, \bibinfo {author} {\bibfnamefont {A.}~\bibnamefont {Douglas}},\ and\ \bibinfo {author} {\bibfnamefont {M.~e.~a.}\ \bibnamefont {Szurek}},\ }\bibfield  {title} {\bibinfo {title} {Dipolar quantum solids emerging in a hubbard quantum simulator},\ }\href@noop {} {\bibfield  {journal} {\bibinfo  {journal} {Nature}\ }\textbf {\bibinfo {volume} {622}},\ \bibinfo {pages} {724} (\bibinfo {year} {2023})}\BibitemShut {NoStop}%
\bibitem [{\citenamefont {Zöllner}\ \emph {et~al.}(2011)\citenamefont {Zöllner}, \citenamefont {Bruun}, \citenamefont {Pethick},\ and\ \citenamefont {Reimann}}]{ref3}%
  \BibitemOpen
  \bibfield  {author} {\bibinfo {author} {\bibfnamefont {S.}~\bibnamefont {Zöllner}}, \bibinfo {author} {\bibfnamefont {G.~M.}\ \bibnamefont {Bruun}}, \bibinfo {author} {\bibfnamefont {C.~J.}\ \bibnamefont {Pethick}},\ and\ \bibinfo {author} {\bibfnamefont {S.~M.}\ \bibnamefont {Reimann}},\ }\bibfield  {title} {\bibinfo {title} {Bosonic and fermionic dipoles on a ring},\ }\href@noop {} {\bibfield  {journal} {\bibinfo  {journal} {Phys. Rev. Lett.}\ }\textbf {\bibinfo {volume} {107}},\ \bibinfo {pages} {035301} (\bibinfo {year} {2011})}\BibitemShut {NoStop}%
\bibitem [{\citenamefont {Zöllner}(2011)}]{ref4}%
  \BibitemOpen
  \bibfield  {author} {\bibinfo {author} {\bibfnamefont {S.}~\bibnamefont {Zöllner}},\ }\bibfield  {title} {\bibinfo {title} {Ground states of dipolar gases in quasi-one-dimensional ring traps},\ }\href@noop {} {\bibfield  {journal} {\bibinfo  {journal} {Phys. Rev. A}\ }\textbf {\bibinfo {volume} {84}},\ \bibinfo {pages} {063619} (\bibinfo {year} {2011})}\BibitemShut {NoStop}%
\bibitem [{\citenamefont {Astrakharchik}\ \emph {et~al.}(2008)\citenamefont {Astrakharchik}, \citenamefont {Morigi}, \citenamefont {Chiara},\ and\ \citenamefont {Boronat}}]{ref5}%
  \BibitemOpen
  \bibfield  {author} {\bibinfo {author} {\bibfnamefont {G.~E.}\ \bibnamefont {Astrakharchik}}, \bibinfo {author} {\bibfnamefont {G.}~\bibnamefont {Morigi}}, \bibinfo {author} {\bibfnamefont {G.~D.}\ \bibnamefont {Chiara}},\ and\ \bibinfo {author} {\bibfnamefont {J.}~\bibnamefont {Boronat}},\ }\bibfield  {title} {\bibinfo {title} {Ground state of low-dimensional dipolar gases: Linear and zigzag chains},\ }\href@noop {} {\bibfield  {journal} {\bibinfo  {journal} {Phys. Rev. A}\ }\textbf {\bibinfo {volume} {78}},\ \bibinfo {pages} {063622} (\bibinfo {year} {2008})}\BibitemShut {NoStop}%
\bibitem [{\citenamefont {Bera}\ \emph {et~al.}(2019)\citenamefont {Bera}, \citenamefont {Chakrabarti}, \citenamefont {Gammal}, \citenamefont {Tsatsos}, \citenamefont {Lekala}, \citenamefont {Chatterjee}, \citenamefont {L{\'e}v{\^e}que},\ and\ \citenamefont {Lode}}]{sangita_sci.rep}%
  \BibitemOpen
  \bibfield  {author} {\bibinfo {author} {\bibfnamefont {S.}~\bibnamefont {Bera}}, \bibinfo {author} {\bibfnamefont {B.}~\bibnamefont {Chakrabarti}}, \bibinfo {author} {\bibfnamefont {A.}~\bibnamefont {Gammal}}, \bibinfo {author} {\bibfnamefont {M.~C.}\ \bibnamefont {Tsatsos}}, \bibinfo {author} {\bibfnamefont {M.~L.}\ \bibnamefont {Lekala}}, \bibinfo {author} {\bibfnamefont {B.}~\bibnamefont {Chatterjee}}, \bibinfo {author} {\bibfnamefont {C.}~\bibnamefont {L{\'e}v{\^e}que}},\ and\ \bibinfo {author} {\bibfnamefont {A.~U.~J.}\ \bibnamefont {Lode}},\ }\bibfield  {title} {\bibinfo {title} {Sorting fermionization from crystallization in many-boson wavefunctions},\ }\href@noop {} {\bibfield  {journal} {\bibinfo  {journal} {Scientific Reports}\ }\textbf {\bibinfo {volume} {9}},\ \bibinfo {pages} {17873} (\bibinfo {year} {2019})}\BibitemShut {NoStop}%
\bibitem [{\citenamefont {Streltsov}\ \emph {et~al.}(2006)\citenamefont {Streltsov}, \citenamefont {Alon},\ and\ \citenamefont {Cederbaum}}]{Streltsov:2006}%
  \BibitemOpen
  \bibfield  {author} {\bibinfo {author} {\bibfnamefont {A.~I.}\ \bibnamefont {Streltsov}}, \bibinfo {author} {\bibfnamefont {O.~E.}\ \bibnamefont {Alon}},\ and\ \bibinfo {author} {\bibfnamefont {L.~S.}\ \bibnamefont {Cederbaum}},\ }\bibfield  {title} {\bibinfo {title} {General variational many-body theory with complete self-consistency for trapped bosonic systems},\ }\href {https://doi.org/10.1103/PhysRevA.73.063626} {\bibfield  {journal} {\bibinfo  {journal} {Phys. Rev. A}\ }\textbf {\bibinfo {volume} {73}},\ \bibinfo {pages} {063626} (\bibinfo {year} {2006})}\BibitemShut {NoStop}%
\bibitem [{\citenamefont {Streltsov}\ \emph {et~al.}(2007)\citenamefont {Streltsov}, \citenamefont {Alon},\ and\ \citenamefont {Cederbaum}}]{Streltsov:2007}%
  \BibitemOpen
  \bibfield  {author} {\bibinfo {author} {\bibfnamefont {A.~I.}\ \bibnamefont {Streltsov}}, \bibinfo {author} {\bibfnamefont {O.~E.}\ \bibnamefont {Alon}},\ and\ \bibinfo {author} {\bibfnamefont {L.~S.}\ \bibnamefont {Cederbaum}},\ }\bibfield  {title} {\bibinfo {title} {Role of excited states in the splitting of a trapped interacting bose-einstein condensate by a time-dependent barrier},\ }\href {https://doi.org/10.1103/PhysRevLett.99.030402} {\bibfield  {journal} {\bibinfo  {journal} {Phys. Rev. Lett.}\ }\textbf {\bibinfo {volume} {99}},\ \bibinfo {pages} {030402} (\bibinfo {year} {2007})}\BibitemShut {NoStop}%
\bibitem [{\citenamefont {Alon}\ \emph {et~al.}(2007)\citenamefont {Alon}, \citenamefont {Streltsov},\ and\ \citenamefont {Cederbaum}}]{Alon:2007}%
  \BibitemOpen
  \bibfield  {author} {\bibinfo {author} {\bibfnamefont {O.~E.}\ \bibnamefont {Alon}}, \bibinfo {author} {\bibfnamefont {A.~I.}\ \bibnamefont {Streltsov}},\ and\ \bibinfo {author} {\bibfnamefont {L.~S.}\ \bibnamefont {Cederbaum}},\ }\bibfield  {title} {\bibinfo {title} {Unified view on multiconfigurational time propagation for systems consisting of identical particles},\ }\href {https://doi.org/10.1063/1.2771159} {\bibfield  {journal} {\bibinfo  {journal} {J. Chem. Phys.}\ }\textbf {\bibinfo {volume} {127}},\ \bibinfo {pages} {154103} (\bibinfo {year} {2007})}\BibitemShut {NoStop}%
\bibitem [{\citenamefont {Alon}\ \emph {et~al.}(2008)\citenamefont {Alon}, \citenamefont {Streltsov},\ and\ \citenamefont {Cederbaum}}]{Alon:2008}%
  \BibitemOpen
  \bibfield  {author} {\bibinfo {author} {\bibfnamefont {O.~E.}\ \bibnamefont {Alon}}, \bibinfo {author} {\bibfnamefont {A.~I.}\ \bibnamefont {Streltsov}},\ and\ \bibinfo {author} {\bibfnamefont {L.~S.}\ \bibnamefont {Cederbaum}},\ }\bibfield  {title} {\bibinfo {title} {Multiconfigurational time-dependent hartree method for bosons: Many-body dynamics of bosonic systems},\ }\href {https://doi.org/10.1103/PhysRevA.77.033613} {\bibfield  {journal} {\bibinfo  {journal} {Phys. Rev. A}\ }\textbf {\bibinfo {volume} {77}},\ \bibinfo {pages} {033613} (\bibinfo {year} {2008})}\BibitemShut {NoStop}%
\bibitem [{\citenamefont {Lode}(2016)}]{Lode:2016}%
  \BibitemOpen
  \bibfield  {author} {\bibinfo {author} {\bibfnamefont {A.~U.~J.}\ \bibnamefont {Lode}},\ }\bibfield  {title} {\bibinfo {title} {Multiconfigurational time-dependent hartree method for bosons with internal degrees of freedom: Theory and composite fragmentation of multicomponent bose-einstein condensates},\ }\href {https://doi.org/10.1103/PhysRevA.93.063601} {\bibfield  {journal} {\bibinfo  {journal} {Phys. Rev. A}\ }\textbf {\bibinfo {volume} {93}},\ \bibinfo {pages} {063601} (\bibinfo {year} {2016})}\BibitemShut {NoStop}%
\bibitem [{\citenamefont {Fasshauer}\ and\ \citenamefont {Lode}(2016)}]{Fasshauer:2016}%
  \BibitemOpen
  \bibfield  {author} {\bibinfo {author} {\bibfnamefont {E.}~\bibnamefont {Fasshauer}}\ and\ \bibinfo {author} {\bibfnamefont {A.~U.~J.}\ \bibnamefont {Lode}},\ }\bibfield  {title} {\bibinfo {title} {Multiconfigurational time-dependent hartree method for fermions: Implementation, exactness, and few-fermion tunneling to open space},\ }\href {https://doi.org/10.1103/PhysRevA.93.033635} {\bibfield  {journal} {\bibinfo  {journal} {Phys. Rev. A}\ }\textbf {\bibinfo {volume} {93}},\ \bibinfo {pages} {033635} (\bibinfo {year} {2016})}\BibitemShut {NoStop}%
\bibitem [{\citenamefont {Lode}\ \emph {et~al.}(2020)\citenamefont {Lode}, \citenamefont {L{\'e}v{\^e}que}, \citenamefont {Madsen}, \citenamefont {Streltsov},\ and\ \citenamefont {Alon}}]{Lode:2020}%
  \BibitemOpen
  \bibfield  {author} {\bibinfo {author} {\bibfnamefont {A.~U.~J.}\ \bibnamefont {Lode}}, \bibinfo {author} {\bibfnamefont {C.}~\bibnamefont {L{\'e}v{\^e}que}}, \bibinfo {author} {\bibfnamefont {L.~B.}\ \bibnamefont {Madsen}}, \bibinfo {author} {\bibfnamefont {A.~I.}\ \bibnamefont {Streltsov}},\ and\ \bibinfo {author} {\bibfnamefont {O.~E.}\ \bibnamefont {Alon}},\ }\bibfield  {title} {\bibinfo {title} {Colloquium: Multiconfigurational time-dependent hartree approaches for indistinguishable particles},\ }\href {https://doi.org/10.1103/RevModPhys.92.011001} {\bibfield  {journal} {\bibinfo  {journal} {Rev. Mod. Phys}\ }\textbf {\bibinfo {volume} {92}},\ \bibinfo {pages} {011001} (\bibinfo {year} {2020})}\BibitemShut {NoStop}%
\bibitem [{\citenamefont {Lin}\ \emph {et~al.}(2020)\citenamefont {Lin}, \citenamefont {Molignini}, \citenamefont {Papariello}, \citenamefont {Tsatsos}, \citenamefont {L{\'e}v{\^e}que}, \citenamefont {Weiner}, \citenamefont {Fasshauer},\ and\ \citenamefont {Chitra}}]{Lin:2020}%
  \BibitemOpen
  \bibfield  {author} {\bibinfo {author} {\bibfnamefont {R.}~\bibnamefont {Lin}}, \bibinfo {author} {\bibfnamefont {P.}~\bibnamefont {Molignini}}, \bibinfo {author} {\bibfnamefont {L.}~\bibnamefont {Papariello}}, \bibinfo {author} {\bibfnamefont {M.~C.}\ \bibnamefont {Tsatsos}}, \bibinfo {author} {\bibfnamefont {C.}~\bibnamefont {L{\'e}v{\^e}que}}, \bibinfo {author} {\bibfnamefont {S.~E.}\ \bibnamefont {Weiner}}, \bibinfo {author} {\bibfnamefont {E.}~\bibnamefont {Fasshauer}},\ and\ \bibinfo {author} {\bibfnamefont {R.}~\bibnamefont {Chitra}},\ }\bibfield  {title} {\bibinfo {title} {Mctdh-x: The multiconfigurational time-dependent hartree method for indistinguishable particles software},\ }\href {https://doi.org/10.1088/2058-9565/ab788b} {\bibfield  {journal} {\bibinfo  {journal} {Quantum Sci. Technol.}\ }\textbf {\bibinfo {volume} {5}},\ \bibinfo {pages} {024004} (\bibinfo {year} {2020})}\BibitemShut {NoStop}%
\bibitem [{\citenamefont {Lode}\ \emph {et~al.}(2024)\citenamefont {Lode}, \citenamefont {Tsatsos}, \citenamefont {Fasshauer}, \citenamefont {Weiner}, \citenamefont {Lin}, \citenamefont {Papariello}, \citenamefont {Molignini}, \citenamefont {L{\'e}v{\^e}que}, \citenamefont {B{\"u}ttner}, \citenamefont {Xiang}, \citenamefont {Dutta},\ and\ \citenamefont {Bilinskaya}}]{MCTDHX}%
  \BibitemOpen
  \bibfield  {author} {\bibinfo {author} {\bibfnamefont {A.~U.~J.}\ \bibnamefont {Lode}}, \bibinfo {author} {\bibfnamefont {M.~C.}\ \bibnamefont {Tsatsos}}, \bibinfo {author} {\bibfnamefont {E.}~\bibnamefont {Fasshauer}}, \bibinfo {author} {\bibfnamefont {S.~E.}\ \bibnamefont {Weiner}}, \bibinfo {author} {\bibfnamefont {R.}~\bibnamefont {Lin}}, \bibinfo {author} {\bibfnamefont {L.}~\bibnamefont {Papariello}}, \bibinfo {author} {\bibfnamefont {P.}~\bibnamefont {Molignini}}, \bibinfo {author} {\bibfnamefont {C.}~\bibnamefont {L{\'e}v{\^e}que}}, \bibinfo {author} {\bibfnamefont {M.}~\bibnamefont {B{\"u}ttner}}, \bibinfo {author} {\bibfnamefont {J.}~\bibnamefont {Xiang}}, \bibinfo {author} {\bibfnamefont {S.}~\bibnamefont {Dutta}},\ and\ \bibinfo {author} {\bibfnamefont {Y.}~\bibnamefont {Bilinskaya}},\ }\href {http://ultracold.org} {\bibinfo {title} {Mctdh-x: The multiconfigurational time-dependent hartree method for indistinguishable particles software}} (\bibinfo {year} {2024})\BibitemShut {NoStop}%
\bibitem [{\citenamefont {Sinha}\ and\ \citenamefont {Santos}(2007)}]{Santos}%
  \BibitemOpen
  \bibfield  {author} {\bibinfo {author} {\bibfnamefont {S.}~\bibnamefont {Sinha}}\ and\ \bibinfo {author} {\bibfnamefont {L.}~\bibnamefont {Santos}},\ }\bibfield  {title} {\bibinfo {title} {Cold dipolar gases in quasi-one-dimensional geometries},\ }\href@noop {} {\bibfield  {journal} {\bibinfo  {journal} {Phys. Rev. Lett.}\ }\textbf {\bibinfo {volume} {99}},\ \bibinfo {pages} {140406} (\bibinfo {year} {2007})}\BibitemShut {NoStop}%
\bibitem [{\citenamefont {Deuretzbacher}\ \emph {et~al.}(2010)\citenamefont {Deuretzbacher}, \citenamefont {Cremon},\ and\ \citenamefont {Reimann}}]{Deuretzbacher}%
  \BibitemOpen
  \bibfield  {author} {\bibinfo {author} {\bibfnamefont {F.}~\bibnamefont {Deuretzbacher}}, \bibinfo {author} {\bibfnamefont {J.~C.}\ \bibnamefont {Cremon}},\ and\ \bibinfo {author} {\bibfnamefont {S.~M.}\ \bibnamefont {Reimann}},\ }\bibfield  {title} {\bibinfo {title} {Ground-state properties of few dipolar bosons in a quasi-one-dimensional harmonic trap},\ }\href@noop {} {\bibfield  {journal} {\bibinfo  {journal} {Phys. Rev. A}\ }\textbf {\bibinfo {volume} {81}},\ \bibinfo {pages} {063616} (\bibinfo {year} {2010})}\BibitemShut {NoStop}%
\bibitem [{\citenamefont {Chatterjee}\ and\ \citenamefont {Lode}(2018)}]{Budhaditya}%
  \BibitemOpen
  \bibfield  {author} {\bibinfo {author} {\bibfnamefont {B.}~\bibnamefont {Chatterjee}}\ and\ \bibinfo {author} {\bibfnamefont {A.~U.~J.}\ \bibnamefont {Lode}},\ }\bibfield  {title} {\bibinfo {title} {Order parameter and detection for a finite ensemble of crystallized one-dimensional dipolar bosons in optical lattices},\ }\href@noop {} {\bibfield  {journal} {\bibinfo  {journal} {Phys. Rev. A}\ }\textbf {\bibinfo {volume} {98}},\ \bibinfo {pages} {053624} (\bibinfo {year} {2018})}\BibitemShut {NoStop}%
\bibitem [{\citenamefont {Molignini}\ \emph {et~al.}(2022)\citenamefont {Molignini}, \citenamefont {Lévêque}, \citenamefont {Keßler}, \citenamefont {Jaksch}, \citenamefont {Chitra},\ and\ \citenamefont {Lode}}]{Molignini:2022}%
  \BibitemOpen
  \bibfield  {author} {\bibinfo {author} {\bibfnamefont {P.}~\bibnamefont {Molignini}}, \bibinfo {author} {\bibfnamefont {C.}~\bibnamefont {Lévêque}}, \bibinfo {author} {\bibfnamefont {H.}~\bibnamefont {Keßler}}, \bibinfo {author} {\bibfnamefont {D.}~\bibnamefont {Jaksch}}, \bibinfo {author} {\bibfnamefont {R.}~\bibnamefont {Chitra}},\ and\ \bibinfo {author} {\bibfnamefont {A.~U.~J.}\ \bibnamefont {Lode}},\ }\bibfield  {title} {\bibinfo {title} {Crystallization via cavity-assisted infinite-range interactions},\ }\href {https://doi.org/10.1103/PhysRevA.106.L011701} {\bibfield  {journal} {\bibinfo  {journal} {Phys. Rev. A}\ }\textbf {\bibinfo {volume} {106}},\ \bibinfo {pages} {L011701} (\bibinfo {year} {2022})}\BibitemShut {NoStop}%
\bibitem [{\citenamefont {Novotny}\ \emph {et~al.}(2023)\citenamefont {Novotny}, \citenamefont {Vijayan1}, \citenamefont {Piotrowski}, \citenamefont {Gonzalez-Ballestero}, \citenamefont {Weber},\ and\ \citenamefont {Romero-Isart}}]{Lukas:2023}%
  \BibitemOpen
  \bibfield  {author} {\bibinfo {author} {\bibfnamefont {L.}~\bibnamefont {Novotny}}, \bibinfo {author} {\bibfnamefont {J.}~\bibnamefont {Vijayan1}}, \bibinfo {author} {\bibfnamefont {J.}~\bibnamefont {Piotrowski}}, \bibinfo {author} {\bibfnamefont {C.}~\bibnamefont {Gonzalez-Ballestero}}, \bibinfo {author} {\bibfnamefont {K.}~\bibnamefont {Weber}},\ and\ \bibinfo {author} {\bibfnamefont {O.}~\bibnamefont {Romero-Isart}},\ }\bibfield  {title} {\bibinfo {title} {Cavity-mediated long-range interactions in levitated optomechanics},\ }\href {https://doi.org/10.21203/rs.3.rs-3310148/v1} {\  (\bibinfo {year} {2023})}\BibitemShut {NoStop}%
\bibitem [{\citenamefont {Kramer}\ and\ \citenamefont {Saraceno}(1981)}]{TDVM81}%
  \BibitemOpen
  \bibfield  {author} {\bibinfo {author} {\bibfnamefont {P.}~\bibnamefont {Kramer}}\ and\ \bibinfo {author} {\bibfnamefont {M.}~\bibnamefont {Saraceno}},\ }\href@noop {} {\emph {\bibinfo {title} {Geometry of the Time-Dependent Variational Principle in Quantum Mechanics}}},\ \bibinfo {series} {Lecture Notes in Physics}, Vol.\ \bibinfo {volume} {140}\ (\bibinfo  {publisher} {Springer},\ \bibinfo {year} {1981})\BibitemShut {NoStop}%
\bibitem [{\citenamefont {Kvaal}(2013)}]{variational1}%
  \BibitemOpen
  \bibfield  {author} {\bibinfo {author} {\bibfnamefont {S.}~\bibnamefont {Kvaal}},\ }\bibfield  {title} {\bibinfo {title} {Variational formulations of the coupled-cluster method in quantum chemistry},\ }\href@noop {} {\bibfield  {journal} {\bibinfo  {journal} {Molecular Physics}\ }\textbf {\bibinfo {volume} {111}},\ \bibinfo {pages} {1100} (\bibinfo {year} {2013})}\BibitemShut {NoStop}%
\bibitem [{\citenamefont {McLachlan}(1964)}]{variational3}%
  \BibitemOpen
  \bibfield  {author} {\bibinfo {author} {\bibfnamefont {A.}~\bibnamefont {McLachlan}},\ }\bibfield  {title} {\bibinfo {title} {A variational solution of the time-dependent schrodinger equation},\ }\href@noop {} {\bibfield  {journal} {\bibinfo  {journal} {Molecular Physics}\ }\textbf {\bibinfo {volume} {8}},\ \bibinfo {pages} {39} (\bibinfo {year} {1964})}\BibitemShut {NoStop}%
\bibitem [{\citenamefont {Cao}\ \emph {et~al.}(2017)\citenamefont {Cao}, \citenamefont {Bolsinger}, \citenamefont {Mistakidis}, \citenamefont {Koutentakis}, \citenamefont {Krönke}, \citenamefont {Schurer},\ and\ \citenamefont {Schmelcher}}]{variational4}%
  \BibitemOpen
  \bibfield  {author} {\bibinfo {author} {\bibfnamefont {L.}~\bibnamefont {Cao}}, \bibinfo {author} {\bibfnamefont {V.}~\bibnamefont {Bolsinger}}, \bibinfo {author} {\bibfnamefont {S.~I.}\ \bibnamefont {Mistakidis}}, \bibinfo {author} {\bibfnamefont {G.~M.}\ \bibnamefont {Koutentakis}}, \bibinfo {author} {\bibfnamefont {S.}~\bibnamefont {Krönke}}, \bibinfo {author} {\bibfnamefont {J.~M.}\ \bibnamefont {Schurer}},\ and\ \bibinfo {author} {\bibfnamefont {P.}~\bibnamefont {Schmelcher}},\ }\bibfield  {title} {\bibinfo {title} {A unified ab initio approach to the correlated quantum dynamics of ultracold fermionic and bosonic mixtures},\ }\href@noop {} {\bibfield  {journal} {\bibinfo  {journal} {J. Chem. Phys.}\ }\textbf {\bibinfo {volume} {147}},\ \bibinfo {pages} {044106} (\bibinfo {year} {2017})}\BibitemShut {NoStop}%
\end{thebibliography}
%

\end{document}